\documentclass[journal]{IEEEtran}
\usepackage{cite}
\usepackage{multirow}
\usepackage{hyperref}  
\usepackage[caption=false,font=normalsize,labelfont=sf,textfont=sf]{subfig}
\ifCLASSINFOpdf
  \usepackage[pdftex]{graphicx}
  \graphicspath{{figures/}}
  \DeclareGraphicsExtensions{.pdf,.jpeg,.png}
\else

\fi

\usepackage{stfloats}
\usepackage{graphicx}

\usepackage{amsmath}

\usepackage{url}
\begin{document}

\title{dCoNNear: An Artifact-Free Neural Network Architecture for Closed-loop Audio Signal Processing}
\author{Chuan~Wen,
        Guy~Torfs,~\IEEEmembership{Senior Member,~IEEE},Sarah~Verhulst,~\IEEEmembership{Senior Member,~IEEE}
\thanks{Chuan Wen and Sarah Verhulst are with the Hearing Technology Lab, Department of Information Technology, Ghent University, Ghent, Belgium.\\
Guy Torfs is with IDLAB, Department of Information Technology, Ghent University, Ghent, Belgium.\\
E-mail: chuan.wen@ugent.be; guy.torfs@ugent.be; sarah.verhulst@ugent.be (Corresponding author: Chuan Wen) \\
This work was supported by  FWO Machine Hearing 2.0 (216318G) and EIC-Transition EarDiTech (101058278).
}
}



\maketitle
\begin{abstract}
Recent advances in deep neural networks (DNNs) have significantly improved various audio processing applications, including speech enhancement, synthesis, and hearing-aid algorithms.
DNN-based closed-loop systems have gained popularity in these applications due to their robust performance and ability to adapt to diverse conditions. Despite their effectiveness, current DNN-based closed-loop systems often suffer from sound quality degradation caused by artifacts introduced by suboptimal sampling methods. To address this challenge, we introduce dCoNNear, a novel DNN architecture designed for seamless integration into closed-loop frameworks. This architecture specifically aims to prevent the generation of spurious artifacts-most notably tonal and aliasing artifacts arising from non-ideal sampling layers.
We demonstrate the effectiveness of dCoNNear through a proof-of-principle example within a closed-loop framework that employs biophysically realistic models of auditory processing for both normal and hearing-impaired profiles to design personalized hearing-aid algorithms.
We further validate the broader applicability and artifact-free performance of dCoNNear through speech-enhancement experiments, confirming its ability to improve perceptual sound quality without introducing architecture-induced artifacts. 
Our results show that dCoNNear not only accurately simulates all processing stages of existing non-DNN biophysical models but also significantly improves sound quality by eliminating audible artifacts in both hearing-aid and speech-enhancement applications. This study offers a robust, perceptually transparent closed-loop processing framework for high-fidelity audio applications.
\end{abstract}


%
\IEEEpeerreviewmaketitle

\section{Introduction}
\begin{figure}[t]
\centering
\includegraphics[width=0.35\textwidth]{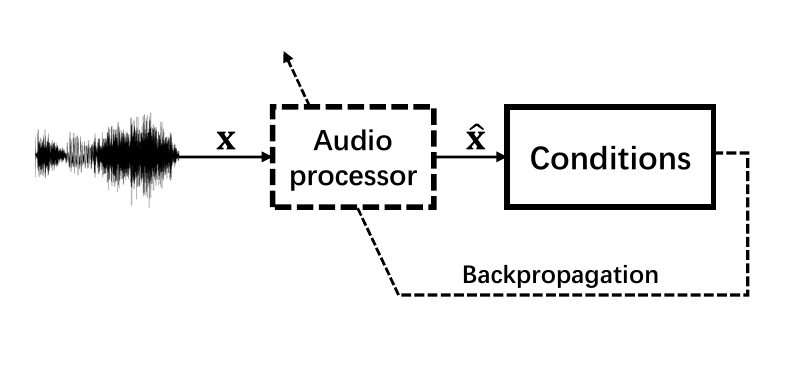}
\caption{Block diagram of the closed-loop framework for audio applications.}
\label{fig_closed}
\vspace{-10pt}
\end{figure}

Recent advances in deep neural networks (DNN) have demonstrated remarkable success in various audio processing applications, such as speech synthesis \cite{donahue2018adversarial,kong2020hifi,bak2023avocodo}, speech recognition \cite{abdel2014convolutional}, speech enhancement \cite{park2017ConvSE}, hearing-aid algorithms \cite{drakopoulos2023neural,drakopoulos2023dnn,drakopoulos2022differentiable}, and others. DNN-based closed-loop frameworks have become increasingly prevalent across various audio applications, primarily due to their notable performance improvements and high flexibility in adapting to diverse conditions.   
The generalized framework illustrated in Fig. \ref{fig_closed} integrates an audio processor with a condition module that can be customized for different tasks. These include using generative adversarial networks (GAN) for audio synthesis \cite{donahue2018adversarial,kong2020hifi,bak2023avocodo}, employing DNN-based loss functions for speech enhancement \cite{germain2019speech,SEQualityNet,xu2021deep,fu2022metricgan}, and implementing biophysically inspired closed-loop frameworks for designing individualized hearing-aid algorithms \cite{drakopoulos2023neural,drakopoulos2023dnn,drakopoulos2022differentiable}.

However, current closed-loop frameworks are corrupted by the non-ideal downsampling and upsampling processes between the NN layers. The problematic operations can generate undesired artifacts, which degrade the resulting audio quality \cite{pons2021upsampling}. Specifically, typical downsampling methods, such as strided convolutions and average pooling, often lack low-pass filtering, leading to aliasing artifacts in low-frequency bands \cite{zhang2019making,bak2023avocodo}. Aliasing artifacts resulting from non-ideal downsampling methods degrade the harmonic components in the speech synthesis,  leading to a noticeable decline in perceptual quality \cite{zaidi2022daft}. 
On the other hand, upsampling methods, such as transposed convolutions and pixel convolutions, tend to generate tonal artifacts that introduce noise in the high-frequency bands \cite{pons2021upsampling,bak2023avocodo}. Additionally, upsampling operations can produce imaging artifacts, where low frequencies are mirrored in high-frequency bands due to spectral replicas \cite{bak2023avocodo,shang2023analysis}.
For instance, \cite{donahue2018adversarial} applied GANs to unsupervised audio generation. However, the use of transposed convolutions introduced pitched noise in the synthesized audio samples.
Furthermore, \cite{drakopoulos2023neural} proposed a bio-inspired DNN-based closed-loop framework for designing hearing-aid (HA) algorithms. The autoencoder-based model (CoNNear) that is used within the closed-loop system comprises a differentiable description of the cochlea, inner-hair-cell (IHC), and auditory-nerve fiber (ANF) processing stages. However, the network produces audible artifacts that compromise the sound quality of the HA model. In these autoencoder-based auditory models (CoNNear\textsubscript{cochlear}, CoNNear\textsubscript{IHC} and CoNNear\textsubscript{ANF}), the undesired artifacts originate from the transposed convolutions in the decoder.

To eliminate aliasing during upsampling and downsampling, Gaussian blur and low-pass filters can be used, as adopted in Alias-CNN \cite{zhang2019making} and stylegan3 \cite{karras2021alias}, respectively. However, these methods resulted in reduced performance in speech synthesis tasks \cite{shang2023analysis}. \cite{shang2023analysis} modified the transposed convolution layer to blend high-frequency features from the input with low-pass filtered features at each upsampling stage. It showed potential in maintaining high performance while reducing aliasing artifacts. 
Alternative upsampling methods like linear interpolation have been proposed to eliminate the tonal artifacts \cite{odena2016deconvolution,stoller2018waveUNet}. The linear interpolation can solve the problem of tonal artifacts, but still leaves imaging artifacts in the high-frequency bands \cite{shang2023analysis}. 

We propose dCoNNear, a novel architecture aimed at minimizing tonal and aliasing artifacts that commonly arise from suboptimal downsampling and upsampling operations in closed-loop DNN systems. We illustrate the approach for hearing-aid algorithms \cite{drakopoulos2023neural} with high sound quality. Beyond hearing-aid applications, we demonstrate the effectiveness of dCoNNear in a closed-loop speech enhancement task, confirming its broader applicability to diverse wave-to-wave audio processing tasks.
The dCoNNear is inspired by temporal convolutional networks (TCN) \cite{lea2016temporal} and deep feedforward sequential memory networks (DFSMN) \cite{zhang2018DFSMN}. It comprises a sequence of stacked memory blocks. For each memory block, depthwise dilated 1-D convolutions are employed to model the long-term dependencies of auditory and audio processing (e.g. cochlear impulse response and neuronal adaptation). This design eliminates the need for downsampling and upsampling, addressing limitations in prior CoNNear-based systems. 
The dCoNNear architecture is applied across all auditory processing stages in the closed-loop system, including cochlear, IHC, and ANF, as well as the sound processor for both hearing-aid and speech-enhancement algorithms. We show that dCoNNear accurately simulates all processing stages of non-DNN-based SOTA biophysical auditory processing, without introducing spurious and audible artifacts in the resulting closed-loop trained audio applications.

The paper is organized as follows: Section II provides an in-depth characterization of the artifacts. Section III introduces the proposed artifact-free dCoNNear-based closed-loop system. Section IV details the experimental procedures, while Section V outlines the evaluation methods. Section VI presents the results for the hearing-aid task, while Section VII demonstrates the speech-enhancement application. 
Section VIII discusses the findings, and Section IX presents a conclusion.

\section{Characteristics of upsampling Artifacts}
The upsampling artifacts are well documented, \cite{pons2021upsampling}  has demonstrated that transposed convolutions and subpixel convolutions introduce tonal artifacts, while nearest neighbor interpolation is prone to generating filtering artifacts. Additionally, during the upsampling process, the spectrum is cyclically repeated at the sampling rate, causing low-frequency components to be mirrored into high-frequency bands. This phenomenon leads to the replication of artifacts introduced by problematic upsampling methods, and the unintended presence of low-frequency components in high-frequency bands, commonly referred to as imaging artifacts. However, \cite{pons2021upsampling} analyzed only the artifacts after weight initialization, without providing a systematic analysis of the impact of the learning process on upsampling artifacts.
In this section, we systematically characterize the artifacts generated by transposed convolutions, subpixel convolutions, and nearest neighbor interpolation when training the network for modeling auditory processing. We then studied artifacts associated with transposed convolution within the context of closed-loop, CNN-based hearing-aid algorithms \cite{drakopoulos2023neural}.

\begin{figure*}[tbp]
\centering
\includegraphics[width=0.8\textwidth]{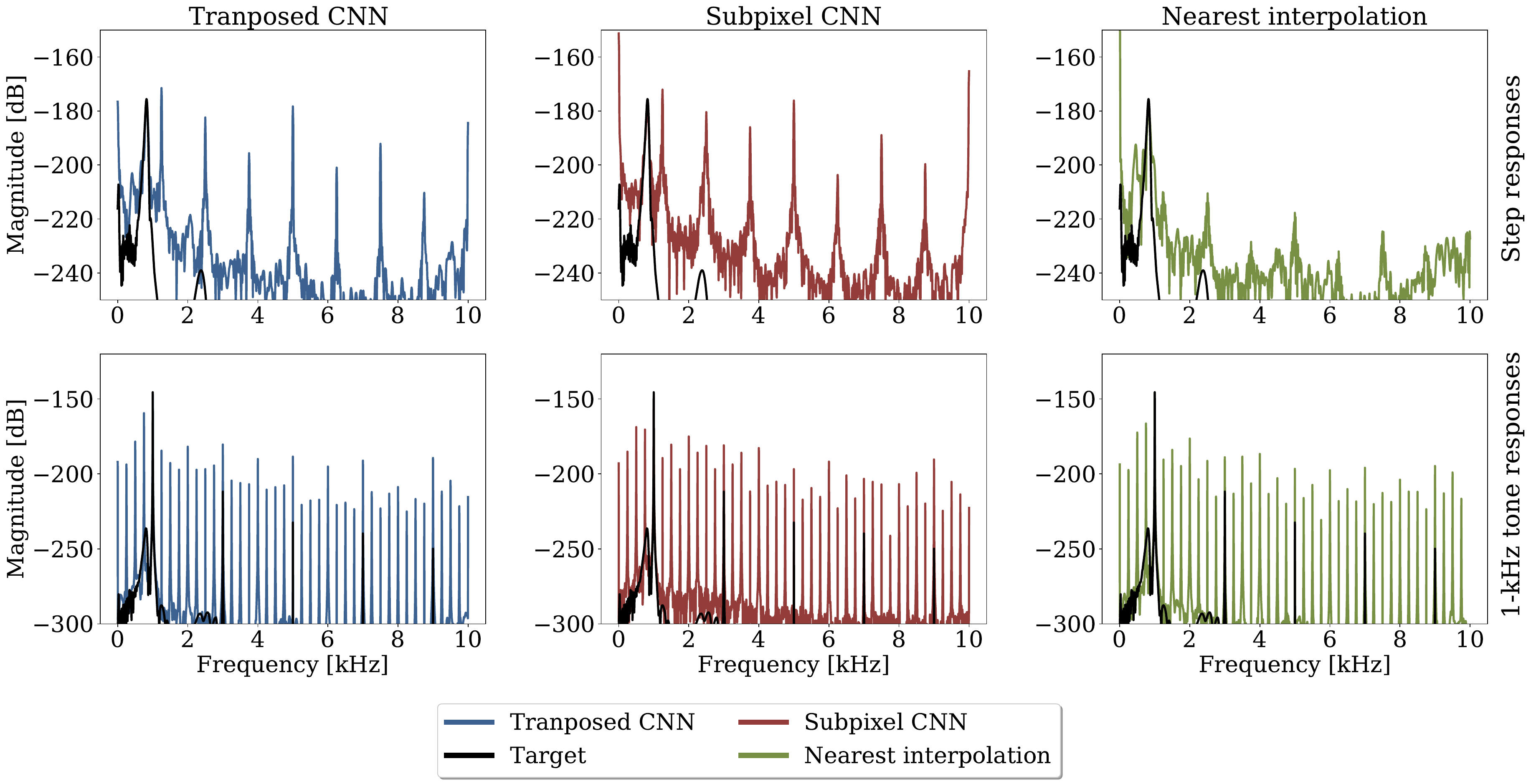}
\caption{Comparing the artifacts between different upsampling methods when training to simulate the TL model. The plots display the magnitude spectrum of BM displacement outputs at a center frequency of 1 kHz. From top to bottom, the stimuli consist of a step input and a 1 kHz pure tone at 70 dB SPL, respectively.}
\label{trained_artifacts}
\end{figure*}

To characterize the upsampling artifacts when employing neural networks to emulate the auditory model, we investigated the artifacts of the autoencoder-based CoNNear\textsubscript{cochlear} \cite{baby2021convolutional} with 4 encoders and decoders, herein referred to as prior CoNNear, between different upsampling methods. CoNNear\textsubscript{cochlear} is a neural network representation of a non-linear transmission-line (TL) model that faithfully simulates the basilar-membrane (BM) displacement of human cochlear processing \cite{verhulst2012TL}. 
We examined three upsampling strategies in the decoders: transposed convolutions, subpixel convolutions, and nearest-neighbor interpolation.
As depicted in Fig. \ref{trained_artifacts}, transposed and subpixel convolutions showed the peaks that were absent in the target model while the nearest-neighbor interpolation showed no obvious extra peaks in the step responses. To mitigate aliasing artifacts introduced by the strided convolution layers, which have a step size of 2, we applied a low-pass filter with a normalized cutoff frequency of 0.5 before each downsampling layer. All upsampling methods showed additional spectral peaks in response to the 1-kHz tone.  
Previous studies have identified periodic tonal artifacts in transposed and subpixel convolutions due to problematic upsampling operators and spectral replicas after weight initialization \cite{pons2021upsampling}. In our findings, these artifacts persisted even after training. 
While nearest-neighbor interpolation is known to avoid tonal artifacts, it failed to effectively suppress imaging artifacts, resulting in noticeable peaks in the 1-kHz tone responses after training.
These results suggest that the training process did not sufficiently resolve the artifacts caused by architectural limitations and spectral replicas, resulting in undesirable distortions in the model's output.

\begin{figure}[tbp]
\centering
\includegraphics[width=0.45\textwidth]{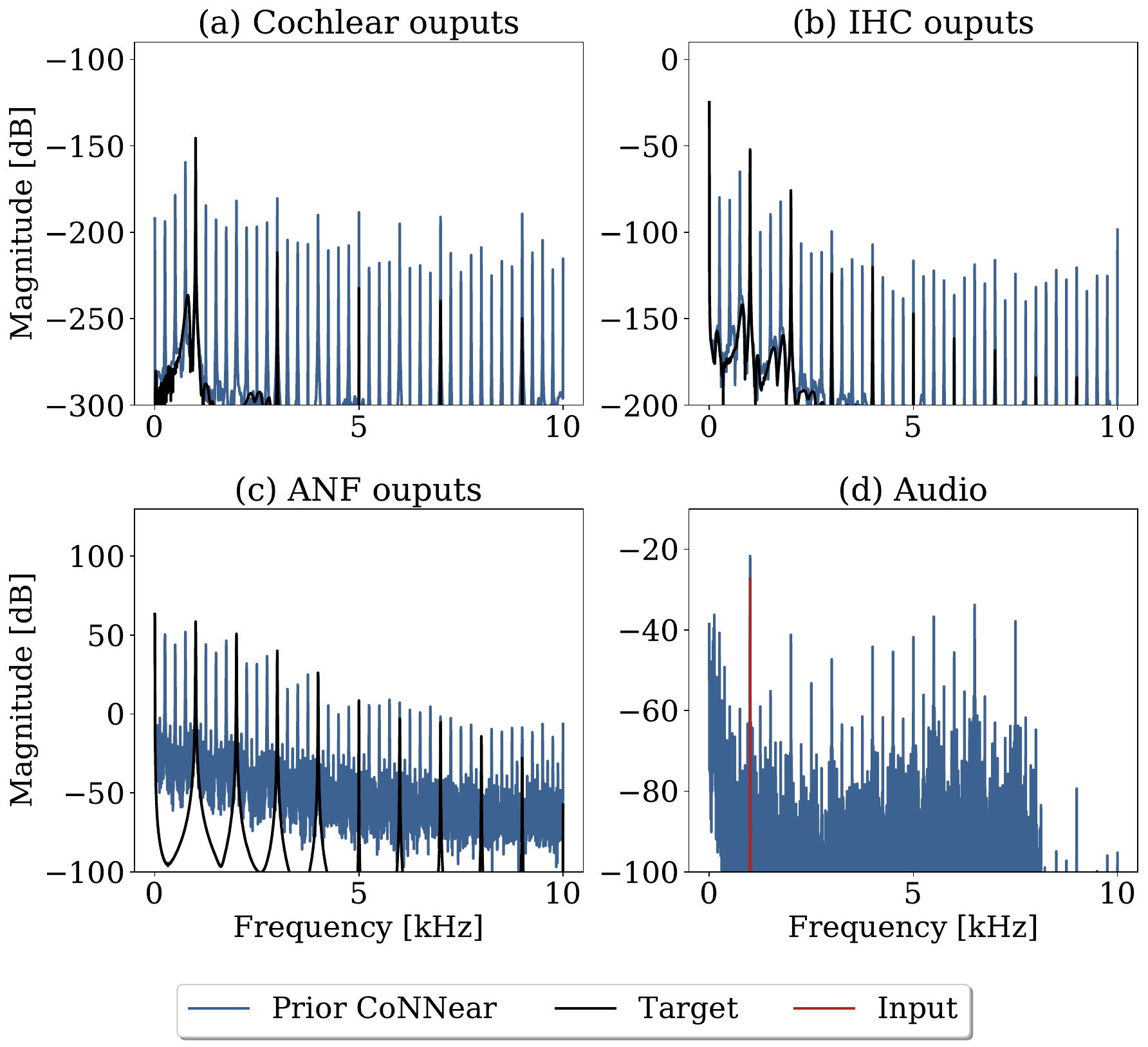}
\caption{The 1kHz tone response at different auditory processing stages compared against the target model for normal-hearing (a-c). (d) The 1-kHz tonal input against the output of the HA model trained from the CoNNear-based framework.}
\label{prior_artifacts_closedLoop}
\end{figure}

The artifacts associated with the autoencoder-based CoNNear models were systematically examined across each auditory processing stage, which include CoNNear\textsubscript{cochlear}, CoNNear\textsubscript{IHC} and CoNNear\textsubscript{ANF} stages as part of the closed-loop system described in \cite{drakopoulos2023neural}.
These artifacts are identified and quantified using the deterministic auditory computational models \cite{verhulst2018computational}, which served as targets during prior CoNNear training. Frequency responses for a 1-kHz tonal input across these stages are illustrated in Fig. \ref{prior_artifacts_closedLoop}(a-c).
Fig. \ref{prior_artifacts_closedLoop}d shows the output of the HA model trained with the prior CoNNear framework to compensate for a high-frequency sloping hearing loss \cite{drakopoulos2023neural}. The prior CoNNear-based models exhibit spectral peaks in the magnitude spectrum of the cochlear, IHC, and ANF responses that are not present in the target model. During the training of the HA model, which minimizes the difference between NH and HI AN responses, the HA models incorporated tonal artifacts as shown in Fig. \ref{prior_artifacts_closedLoop}d. These artifacts degrade the resulting audio quality and should be systematically excluded from the closed-loop framework.

\begin{figure*}[tbp]
\centering
\includegraphics[width=0.8\textwidth,height=\textheight,keepaspectratio]{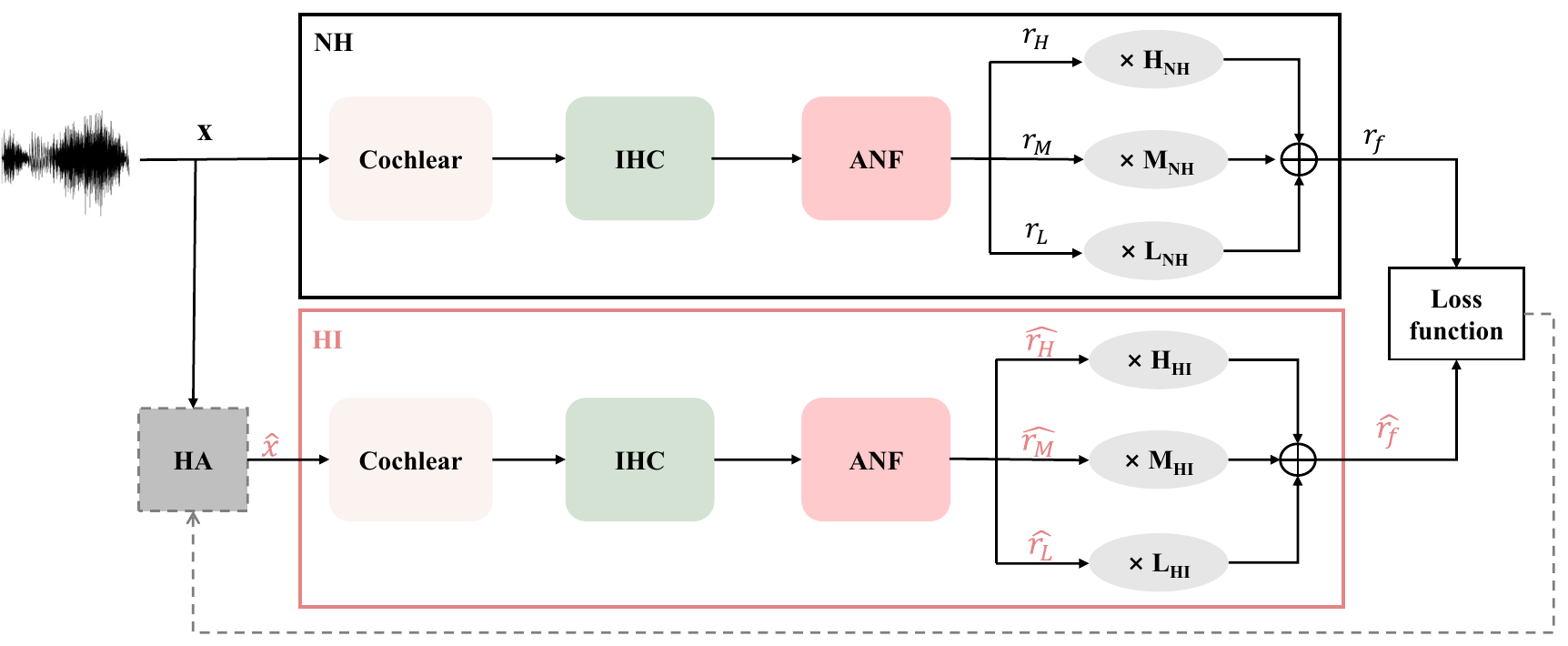}
\caption{Generic diagram of the closed-loop framework for designing individualized hearing-aid algorithms. The three auditory modules—Cochlear, IHC, and ANF—are implemented using deep neural network architectures (e.g., CoNNear and dCoNNear).}
\label{HA_closedLoop}
\end{figure*}


\section{Artifact-free dCoNNear-based closed-loop system}
The artifact-free dCoNNear-based closed-loop framework for individualized DNN-based hearing-aid (HA) model training is illustrated in Fig. \ref{HA_closedLoop}. This framework features two pathways: one for the AN response ($r_f$) of a normal hearing (NH) system, and the other for the response ($\hat{r_f}$) of a hearing-impaired (HI) system.
Each pathway includes three primary components corresponding to distinct auditory processing stages, the cochlea, inner hair cells (IHC), and auditory nerve fibers (ANF). The NH AN response ($r_f$) is generated by combining three types of fibers (high-spontaneous rate (HSR), medium-spontaneous rate (MSR), and low-spontaneous rate (LSR)). These fibers are weighted by $H_{NH}$, $M_{NH}$, and $L_{NH}$, reflecting the typical innervation of each IHC in a healthy auditory system. To simulate hearing-impaired profiles, we adjust these weights to $H_{HI}$, $M_{HI}$, and $L_{HI}$ to reflect the effects of cochlear synaptopathy (CS).
The resulting AN responses ($r_f$) and ($\hat{r_f}$) provide biophysically realistic time-frequency representations of sound (neurograms), simulated at different cochlear locations. These neurograms reflect instantaneous firing rates across cochlear channels with center frequencies (CFs) ranging from 112 Hz to 12 kHz \cite{greenwood1990cochlear}.
To tailor the hearing-impaired periphery to an individual's sensorineural hearing loss (SNHL) profile, adjustments are made using audiometry to simulate the outer-hair-cell (OHC) damage in the cochlea or using auditory evoked potentials \cite{keshishzadeh2021towards} to estimate CS. This involves introducing frequency-dependent OHC loss in CoNNear\textsubscript{cochlear} and/or CS in the CoNNear\textsubscript{ANF} model. The DNN-based HA models are subsequently trained by minimizing a predefined loss function between the simulated NH and HI responses \cite{drakopoulos2023neural}. All auditory elements and HA models within this framework are built upon the dCoNNear topology.

\subsection{dCoNNear Architecture}
\begin{figure*}[t!]
\centering
\includegraphics[width=0.8\textwidth,height=\textheight,keepaspectratio]{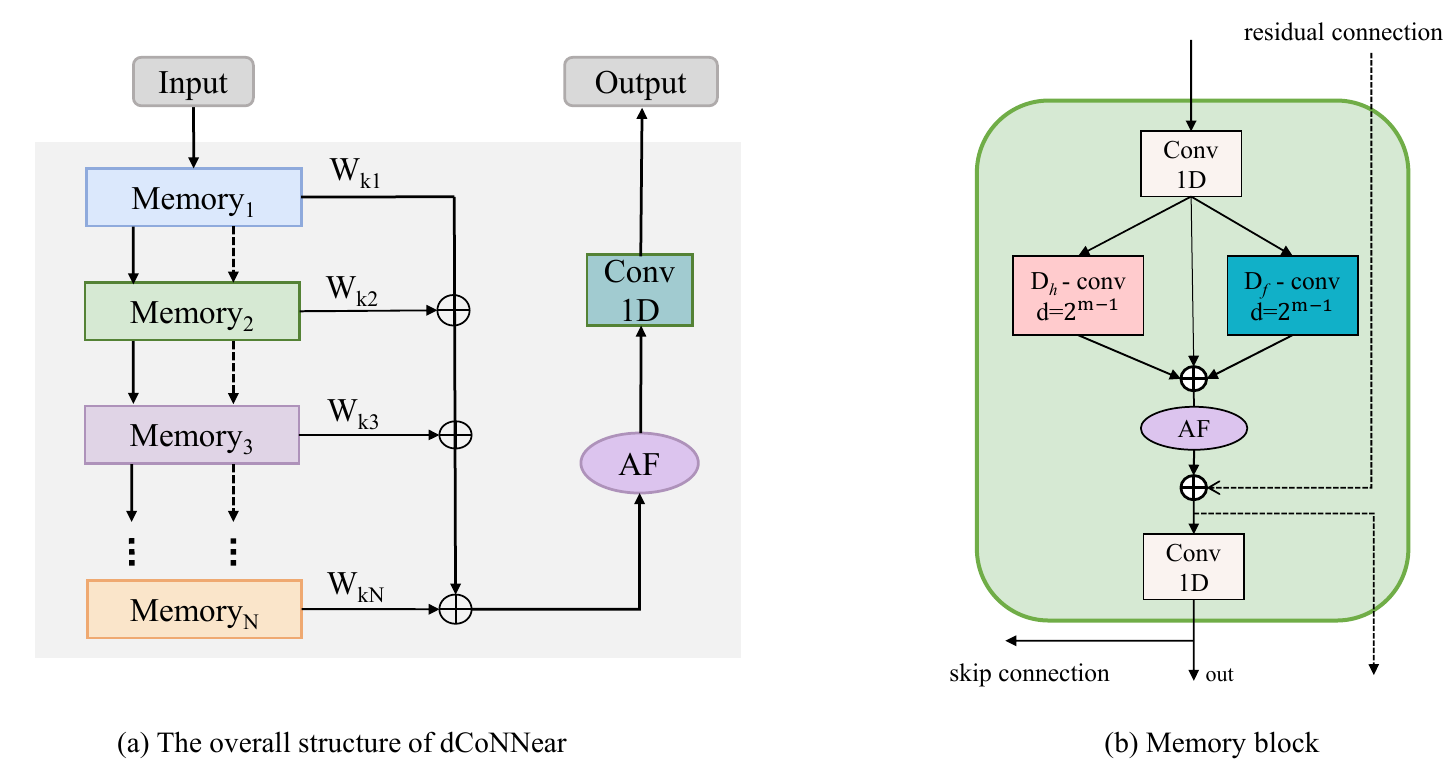}
\caption{(a) The block diagram of the dCoNNear. (b) The diagram of the memory block.}
\label{dCoNNear}
\end{figure*}

To overcome the limitations of conventional encoder-decoder architectures, we introduce the dCoNNear model (Fig. \ref{dCoNNear}), which explicitly addresses structural artifacts, such as aliasing, tonal distortions, and imaging effects, commonly introduced by suboptimal downsampling and upsampling methods \cite{pons2021upsampling, zhang2019making, shang2023analysis}. 
To eliminate the architectural causes of these artifacts, dCoNNear replaces the encoder-decoder structure with a fully convolutional stack of FIR-like memory blocks as illustrated in Fig. \ref{dCoNNear} (b). Each block employs dilated depthwise convolutions to capture long-term temporal dependencies without altering the temporal resolution. By avoiding any explicit downsampling or upsampling operations, the architecture removes the structural conditions under which aliasing and spectral folding arise, thereby preventing the emergence of these artifacts in the output signal. This architecture maintains the ability to capture long-range dependencies while substantially reducing artifacts.

The overall architecture, shown in Fig. \ref{dCoNNear}(a), consists of $M$ memory blocks with dilation factors $2^{m-1}$, repeated R times, resulting in a total of N = M * R blocks. Residual connections between memory blocks facilitate learning by addressing the vanishing gradient problem. Skip connections aggregate the outputs from each memory block through weighted summation, and these aggregated outputs are then processed by a non-linear activation function followed by a convolutional layer to produce the final model outputs \cite{conv-tasnet}.

Each memory block (Fig. \ref{dCoNNear}(b)) begins with a pointwise convolution, followed by two dilated depthwise convolutions—D\textsubscript{\textit{h}}-conv and D\textsubscript{\textit{f}}-conv—which capture historical and future temporal context, respectively. Depthwise convolution applies a single filter per input channel, reducing the number of parameters and helping capture long-term dependencies while maintaining a compact model size. The outputs from the pointwise and depthwise convolution layers are combined, passed through an activation function, and directed to subsequent blocks and skip connection paths. The  learnable FIR-like memory block \cite{zhang2018DFSMN} is formulated as:
\begin{gather}
\tilde{\mathbf{Y}_t^{\ell}} = \mathbf{Y}_t^{\ell} + \sum_{i=0}^{K_1^{\ell}} \mathbf{a}_i^{\ell} \odot \mathbf{Y}_{t-d_1 \cdot i}^{\ell} + \sum_{j=1}^{K_2^{\ell}} \mathbf{b}_j^{\ell} \odot \mathbf{Y}_{t+d_2 \cdot j}^{\ell},\\
\mathbf{Y}_t^{\ell+1} = \mathbf{V}^{\ell}f(\tilde{\mathbf{Y}_t^{\ell}})+\mathbf{U}^{\ell}
\label{eq:formula_FIR}
\end{gather}
Here, $\mathbf{Y}_t^{\ell} = \mathbf{W}^{\ell} \mathbf{h}_t^{\ell} + \mathbf{B}^{\ell}$ represents the linear output of the $\ell$-th projection layer. The parameters $K_1^{\ell}$ and $K_2^{\ell}$ define the look-back and lookahead orders, respectively, indicating the range of past and future data points considered. The dilation factors $d_1$ and $d_2$ help capture broader context information. The symbol $\odot$ denotes element-wise multiplication, and $\mathbf{Y}_t^{\ell+1}$ is the output of the $\ell$-th memory block at time $t$.  
The overall receptive field (RF) can be calculated as:
\begin{equation}
\label{eq:formula_RF}
RF = K_1 + K_2 + R \sum_{i=0}^{M} (K_1-1) 2^i + R \sum_{j=0}^{M} (K_2-1) 2^j
\end{equation}
Where $K_1$ and $K_2$ denote the kernel size of D\textsubscript{\textit{h}}-conv and D\textsubscript{\textit{f}}-conv respectively. This comprehensive approach ensures that the dCoNNear model effectively captures and processes long-term dependencies in audio and auditory signals, enhancing its performance for audio applications.

\subsection{Auditory modules and HA model}
Both normal hearing (NH) and hearing-impaired (HI) auditory peripheries, as shown in Fig. \ref{HA_closedLoop}, utilize biophysically inspired CNN-based models that accurately simulate human cochlear, IHC and ANF processing: dCoNNear\textsubscript{cochlear}, dCoNNear\textsubscript{IHC}, and dCoNNear\textsubscript{ANF}. The dCoNNear\textsubscript{cochlear} and dCoNNear\textsubscript{IHC} follow the dCoNNear topology, while dCoNNear\textsubscript{ANF} employs a three-branch structure to predict the different ANF types (HSR, MSR, and LSR).

When selecting hyperparameters, two primary considerations should be addressed \cite{baby2021convolutional,drakopoulos2023dnn}: the receptive field (RF) and the specific activation function. A larger receptive field (RF) ensures that the CNN-based network accurately simulates the auditory model's impulse response. The selected activation functions need to capture the nonlinear characteristics of auditory processing, such as the compressive growth of the BM vibrations in the cochlear model and the rectification-like behavior of the IHCs. 
The framework's hyperparameters are detailed in Table \ref{tab:model}. Specifically, $K_1$ and $K_2$ denote the kernel sizes of the dilated depthwise convolution layers (D\textsubscript{\textit{h}}-conv and D\textsubscript{\textit{f}}-conv) within each memory block. The parameter $H$ indicates the number of channels in the hidden layers. Additionally, $L$, $L_l$, and $L_r$ represent the input length, left context, and right context, respectively.

The dCoNNear\textsubscript{cochlear} transforms acoustic signals into cochlear basilar-membrane vibrations. As described in \cite{baby2021convolutional}, the hyperbolic tangent (Tanh) activation function was used, because it allows the activations to resemble the input-output relations of the auditory model and supports both positive and negative deflections of the basilar membrane.

The dCoNNear\textsubscript{IHC} model emulates the IHCs' function of sensing basilar membrane vibrations and converting them into IHC receptor potential changes. It employs a Tanh nonlinearity in the first stack and a sigmoid function in the second stack. This approach mirrors the compressive characteristics of the IHC input/output functions, with the sigmoid function effectively capturing the IHC receptor potential as a negative voltage difference \cite{drakopoulos2023neural}.

The dCoNNear\textsubscript{ANF} model converts IHC receptor potentials into firing rates for three types of ANFs. The initial M blocks are shared layers, with each branch containing M blocks to independently generate HSR, MSR, and LSR fiber responses. The overall AN response ($r_f$) is derived by summing the three ANF responses. The ReLU activation function is applied here, as ANF processing differs from cochlear and IHC processing in that it only involves positive output values, without the compressive properties observed in the previous stages. The model features substantial left context windows of 7936 samples, ensuring accurate capture of the time-dependent recovery properties of the ANF models \cite{drakopoulos2021convolutional}.

The DNN-based hearing-aid (HA) model is trained to restore NH AN responses. The Tanh nonlinearity is used to account for the compressive amplification characteristics of the human ear, similar to the wide-dynamic-range-compression (WDRC) strategy used in traditional hearing aids, which automatically adjusts amplification to amplify quiet sounds and compress loud sounds.

\begin{table*}[tbp]
\caption{Hyperparamters of the dCoNNears and HA-model. M and R represent the number of memory blocks and the repetition number; K\textsubscript{1} and K\textsubscript{2} denote the kernel size of D\textsubscript{\textit{h}}-conv and D\textsubscript{\textit{f}}-conv in each memory block; H indicates the channel number of hidden layers; L\textsubscript{l}, and L\textsubscript{r} represent the left context and right context, respectively}
\centering
\begin{tabular}{cccccccccc}
\hline
Models            & M & R & K\textsubscript{1} & K\textsubscript{2} & H   & Activation   & Parameters & $L_l$   & $L_r$   \\ \hline
dCoNNear\textsubscript{cochlear} & 6& 2& 80& 0 &256& Tanh& 1.5M       & 256  & 256  \\ \hline
dCoNNear\textsubscript{IHC}     & 4  & 2  & 32 &32 & 128 & Tanh& 0.3M       & 256  & 256  \\ \hline
dCoNNear\textsubscript{ANF}    & 8   & 2  & 16 &16 & 32  & ReLU         & 0.1M     & 7936 & 256  \\ \hline
HA-model       & 6 & 2 & 32 &  32 & 256 & Tanh& 1.6M       & 7936    & 256    \\ \hline
\end{tabular}
\label{tab:model}

\end{table*}

\subsection{Individualisation of hearing impairment}
To individualise hearing impaired pathway as shown in Fig. \ref{HA_closedLoop}, we adjusted the normal hearing (NH) auditory modules to simulate different degrees of outer hair cell (OHC) loss and cochlear synaptopathy (CS) in the HI pathway.
To simulate OHC loss, we retrained the NH cochlear model using transfer learning, based on a specific gain-loss profile or an individual audiogram \cite{van2020hearing}. The NH auditory nerve (AN) response ($r_f$) is computed as a weighted sum of the three ANF responses ($H_{NH}$ = 13, $M_{NH}$ = 3, and $L_{NH}$ = 3). To simulate CS, the model was modified with adjusted weights $H_{HI}$, $M_{HI}$, and $L_{HI}$ \cite{drakopoulos2023neural}. These personalized adjustments ensured the models accurately represented individual hearing impairments, providing tailored audio processing solutions that minimized artifacts and preserved fidelity.

\section{Training strategy}
The training procedure involved two main stages: (1) training the auditory modules (dCoNNear\textsubscript{cochlear}, dCoNNear\textsubscript{IHC}, and dCoNNear\textsubscript{ANF}), and (2) training the HA model based on the closed-loop framework including the NH and HI pathways. Firstly, we trained the auditory modules using simulated responses from analytical models as targets, following the procedure outlined in \cite{baby2021convolutional,drakopoulos2021convolutional}. Once the auditory elements were trained, their parameters were frozen, and the HA models were then optimized within the closed-loop system as first described in \cite{drakopoulos2023neural}.

\subsection{Training auditory elements}
The auditory module training targets were derived from various analytical models: a biophysical transmission line (TL) model for dCoNNear\textsubscript{cochlear} \cite{verhulst2018computational}, an analytical Hodgkin–Huxley-type model for dCoNNear\textsubscript{IHC} \cite{altoe2017ihc}, and a three-store diffusion model of the ANF synapse for dCoNNear\textsubscript{ANF} \cite{altoe2018ANF}. 
To train the models, we randomly selected 2,310 utterances from the full TIMIT speech corpus \cite{garofolo1993TIMIT} without replacement. The resulting speaker distribution was approximately 70\% male and 30\% female, consistent with the overall gender distribution in TIMIT. The input signals were first upsampled to 100 kHz to accurately solve the reference models, and then downsampled to 20 kHz for training. The root-mean-square (RMS) energy of each sentence was adjusted to a specified sound pressure level (SPL) using the following equation:
\begin{equation}
    signal = p_0 \cdot 10 ^{L_D/20} \cdot signal / RMS (signal)
\label{eq:pressue_level}
\end{equation}
where $p_0$ is the reference pressure of $2 \cdot 10^{-5}$ Pa, and $L_D$ indicates the sound pressure level of the sound measured in decibels (dB).

\subsubsection{Training dCoNNear\textsubscript{cochlear}}
The dCoNNear\textsubscript{cochlear} model was trained to simulate basilar membrane (BM) displacements for 201 center frequencies between 112 Hz and 12 kHz \cite{verhulst2018computational}, as human hearing sensitivity diminishes above 12 kHz. The RMS energy of each sentence was adjusted to $L_D = 70$ dB SPL. The input signal and TL-model outputs were segmented into windows of 2,048 samples, with left ($L_l$) and right ($L_r$) contexts of 256 samples each, resulting in a total input length of 2,560 samples. To facilitate training, a scaling factor of $10^6$ was applied to the simulated BM displacements (expressed in [$\mu m$]), ensuring that the datasets maintained a statistical mean near zero and a standard deviation close to one.

\subsubsection{Training dCoNNear\textsubscript{IHC} and dCoNNear\textsubscript{ANF}}
The dCoNNear\textsubscript{IHC} and dCoNNear\textsubscript{ANF} were trained to simulate IHC potentials and ANF firing rates at 201 uniformly distributed center frequencies between 112 Hz and 12 kHz. The RMS energy of half the sentences was adjusted to 70 dB and 130 dB SPL, respectively, to cover a broad range of instantaneous intensities, enabling the models to capture the characteristic input-output and saturation properties of individual IHCs and ANFs.

For dCoNNear\textsubscript{IHC}, the input signals were segmented into 2,048-sample windows with 256 context samples on both sides. For dCoNNear\textsubscript{ANF}, the analytical IHC and ANF outputs were segmented into 8,192-sample windows, with 7,936 context samples before and 256 samples after each window. The simulated IHC potential outputs were multiplied by a factor of 10, expressed in [dV], and a scaling factor of 0.01 was applied to the simulated ANF outputs, expressed in [x100 spikes/s].

The training data for IHC and ANF, with dimension $N\textsubscript{CF} = 201$, were converted into a one-dimensional dataset of 2,310 $\cdot N\textsubscript{CF}$ samples. This approach assumed that the reference IHC and ANF models had CF-independent parameters, while the BM displacements had CF-dependent impulse responses due to cochlear mechanics \cite{drakopoulos2021convolutional}.
 
\subsection{Training HA model}
To train the HA model, the RMS energy of each utterance was adjusted to 70 dB SPL. The training dataset comprised 2,310 randomly selected utterances from the TIMIT corpus, segmented into windows of 8,192 samples. The inputs included 7,936 left context samples and 256 right context samples to meet the context window requirements of the dCoNNear\textsubscript{ANF} model.
The trained dCoNNear-based auditory models simulated cochlear responses at 201 center frequencies, ranging from 112 Hz to 12 kHz. For training the HA model, 21 equally spaced frequency channels were selected from the 201 to expedite the training process. During HA model optimization, the parameters of the dCoNNear auditory elements were kept frozen, and a predefined loss function was used to minimize the difference between the dCoNNear-simulated NH and HI AN responses.

\subsection{Loss function}
The training strategies from \cite{baby2021convolutional} and \cite{drakopoulos2021convolutional} were employed to train the dCoNNear\textsubscript{cochlear}, dCoNNear\textsubscript{IHC}, and dCoNNear\textsubscript{ANF} models. The loss function for these models was optimized using the mean absolute error (MAE) between the analytical outputs and the predicted dCoNNear outputs.

For training the HA model parameters, we utilized a combined loss function that accounted for different representations of the differences between the NH responses $r_f$ and the HI $\hat{r_f}$. As outlined in \cite{drakopoulos2023neural}, additional constraints, such as population responses, were introduced to the training process to minimize features functionally relevant to hearing loss and auditory perception, given the highly nonlinear nature of auditory processing.
The combined loss function incorporated both the AN responses and the AN population responses across all simulated CFs:
\begin{gather}
\ell_{HA}  = \alpha \cdot \text{MSE}\{r_f(n,w),\hat{r_f}(n,w)\} + \beta \cdot \text{MSE}\{p(n),\hat{p}(n)\}, \\
p(n)  = \sum_{w=1}^{N_{CF}} r(n, w),\\
\hat{p}(n)  = \sum_{w=1}^{N_{CF}} \hat{r}(n, w)
\end{gather}
where $\alpha = 30$ and $\beta = 1$ are the weights that balance the contributions of different losses during training. $p(n)$ and $\hat{p}(n)$ are the NH and HI AN population responses respectively, n corresponds to each sample of the AN population responses, and L to the total number of samples. This approach ensured that the model not only minimized the differences in the AN responses at each CF but also captured the overall population response across CFs, providing a more comprehensive and functionally relevant training strategy for the HA model.

\subsection{Hearing impaired elements}
We optimized the HA model to compensate for a specific hearing-impaired (HI) profile ``Slope35-7,0,0" \cite{drakopoulos2023neural}. This profile represents a sloping high-frequency outer hair cell (OHC) loss beginning at 1 kHz, with hearing thresholds of 35 dB HL at 8 kHz, referred to as Slope35. Additionally, there is a complete loss of low spontaneous rate (LSR) and medium spontaneous rate (MSR) auditory nerve fibers (ANFs) (H\textsubscript{HI} = 0 and M\textsubscript{HI}= 0) and approximately 46\% loss of high spontaneous rate (HSR) ANFs (L\textsubscript{HI} = 7) to simulate the effects of age-related cochlear synaptopathy.

\subsection{Training setup}
The learning rates for model training was set to $10^{-4}$. If the validation loss did not decrease over five consecutive epochs, the learning rate was halved to ensure optimal training progression.
All models were trained for 50 epochs using the Adam optimizer \cite{kingma2014adam}. Training was conducted on three NVIDIA A30 GPUs using the PyTorch platform. 

\section{Evaluation}
To evaluate the artifact-free closed-loop system, we first examined whether the artifacts associated with the prior CoNNear architecture were eliminated and examined the sound quality of the resulting HA models. Additionally, we evaluated the biophysical properties of the dCoNNear auditory elements in comparison to the CoNNear architectures and the hearing loss restoration performance of the trained HA models.

\subsection{Evaluating Artifacts}
To determine whether the artifacts caused by the prior CoNNear were eliminated, we analyzed 70-dB SPL 1-kHz tone responses at various auditory processing stages, as well as the HA models trained within the closed-loop system based on both prior CoNNear and dCoNNear architectures. We quantified artifacts using total harmonic distortion (THD) at each step of the closed-loop system \cite{shmilovitz2005THD}. 
As aliased harmonics are generated by the prior CoNNear models in the 1-kHz tone responses as shown in Fig. \ref{trained_artifacts}, we examined harmonics at one-quarter intervals of the fundamental frequency:
\begin{gather}
    \mathrm{THD_{fractional}} = 20\log_{10} \frac{\sqrt{\sum_{k=5}^{\infty} H_{\frac{k}{4}}^2}}{H_1}
\end{gather}
where $H_1$ is the RMS value of the fundamental component, $ H_{k/4}$ indicates the RMS of the fractional harmonics. A lower THD value indicates less harmonic distortion.

The Speech-to-Reverberation Modulation Energy Ratio (SRMR) \cite{falk2010SRMR}, DNSMOS (overall) \cite{reddy2021dnsmos} and Perceptual evaluation of speech quality (PESQ) \cite{pesq2001perceptual} were used to evaluate the sound quality of the resulting HA models trained using both the prior CoNNear and dCoNNear-based frameworks. For all metrics, higher scores indicate higher sound quality. We evaluated sound quality across three audio types: clean speech, noisy speech, and music. The clean speech dataset consisted of 550 utterances from TIMIT and LibriTTS \cite{zen2019libritts}, which were independent of the training set of HA models. The one-hour noisy speech dataset was generated by mixing the clean speech samples and the noise from Freesound \cite{fonseca2017freesound} at SNR of 0 dB. We used environmental noise from the Freesound database, which typically includes sounds such as street noise, sounds of people talking, machinery hum, and ambient indoor sounds. The music dataset consisted of 100 randomly selected 30-second tracks from FMA \cite{fma_dataset}.

\subsection{Evaluating Auditory Elements}
For each auditory model, we employed two evaluation metrics to assess whether the trained models accurately captured the relevant auditory properties.

For the dCoNNear\textsubscript{cochlear}, we simulated the Q\textsubscript{ERB} and basilar membrane (BM) excitation patterns, as described in \cite{baby2021convolutional}. The Q\textsubscript{ERB} metric characterizes the level-dependent cochlear filter properties derived from the BM impulse response and is calculated as follows:
\begin{gather}
\text{Q}_{\text{ERB}} = \frac{\text{CF}}{\text{ERB}}
\end{gather}
where equivalent-rectangular bandwidth (ERB) is determined from the power spectrum of a simulated BM time-domain response to an acoustic click stimulus with condensation clicks of 100$\mu$s duration at levels of 40 and 70 dB SPL.
Cochlear excitation patterns reflect the nonlinear compressive growth of BM responses when the cochlea is stimulated with pure tones at CF corresponding to the cochlear measurement site. We calculated excitation patterns for 201 CF channels in response to pure tones at 0.5, 1, and 2 kHz frequencies, with levels ranging from 10 to 90 dB SPL.

For the dCoNNear\textsubscript{IHC}, IHC excitation patterns and half-wave rectified IHC receptor potentials were measured. 
Similar to cochlear excitation patterns, IHC excitation patterns show a characteristic half-octave basal-ward shift of their maxima as stimulus level increases. We calculated excitation patterns for all 201 simulated IHC receptor potentials in response to pure tones of 0.5, 1, and 2 kHz frequencies and levels between 10 and 90 dB SPL.
The half-wave rectified receptor potential demonstrates the compression feature of IHC mechanical-to-electrical transduction. To measure this, 4 kHz tonal stimuli with levels from 0 to 100 dB SPL were generated, and the IHC responses were then half-wave rectified by subtracting their DC component. The RMS of the rectified responses was computed for each level.

For the dCoNNear\textsubscript{ANF}, rate-level curves and synchrony levels were measured \cite{drakopoulos2023neural}. Rate-level curves evaluate ANF responses to changes in stimulus level. We generated pure-tone stimuli (50-ms duration, 2.5-ms rise/fall ramp) with levels between 0 and 100 dB at frequencies of approximately 1 and 4 kHz, based on the corresponding CFs of the ANF models (1007 and 3972.7 Hz). 
The rate-level functions were derived by computing the average response 10–40 ms after stimulus onset. 
The ANF synchrony level describes the non-monotonic relation between ANF response and the stimulus level. Fully modulated 400-ms long pure tones with a modulation frequency $f_m$ of 100 Hz and carrier frequencies of  3972.7 Hz (henceforth referred to as 4 kHz) were simulated, and the synchrony-level functions were calculated by extracting the magnitude of the $f_m$ component from the Fourier spectrum of the fibers’ firing rate.

\subsection{Evaluating HA models}
To evaluate the restoration performance of the trained HA models, we used normalized root-mean-square error (NRMSE) in \cite{drakopoulos2023neural}, which was computed between the simulated NH and HI AN population responses, normalised to the maximum of the NH response for each sentence:
\begin{gather}
    \text{NRMSE} = \frac{\text{RMSE}}{\max(p(n))},\\
    \text{RMSE} = \sqrt{\frac{1}{L} \sum_{n=1}^L (p(n) - \hat{p}(n))^2}
\end{gather}
NRMSE calculations were performed on 550 randomly selected clean speech utterances from the TIMIT and LibriTTS \cite{zen2019libritts} utterances, which were excluded from the HA training dataset. These utterances had RMS energy levels ranging from 40 to 70 dB SPL, in 10 dB increments.


\begin{figure*}[tbp]
\centering

\subfloat[Cochlear outputs]{
\includegraphics[width=0.25\textwidth]{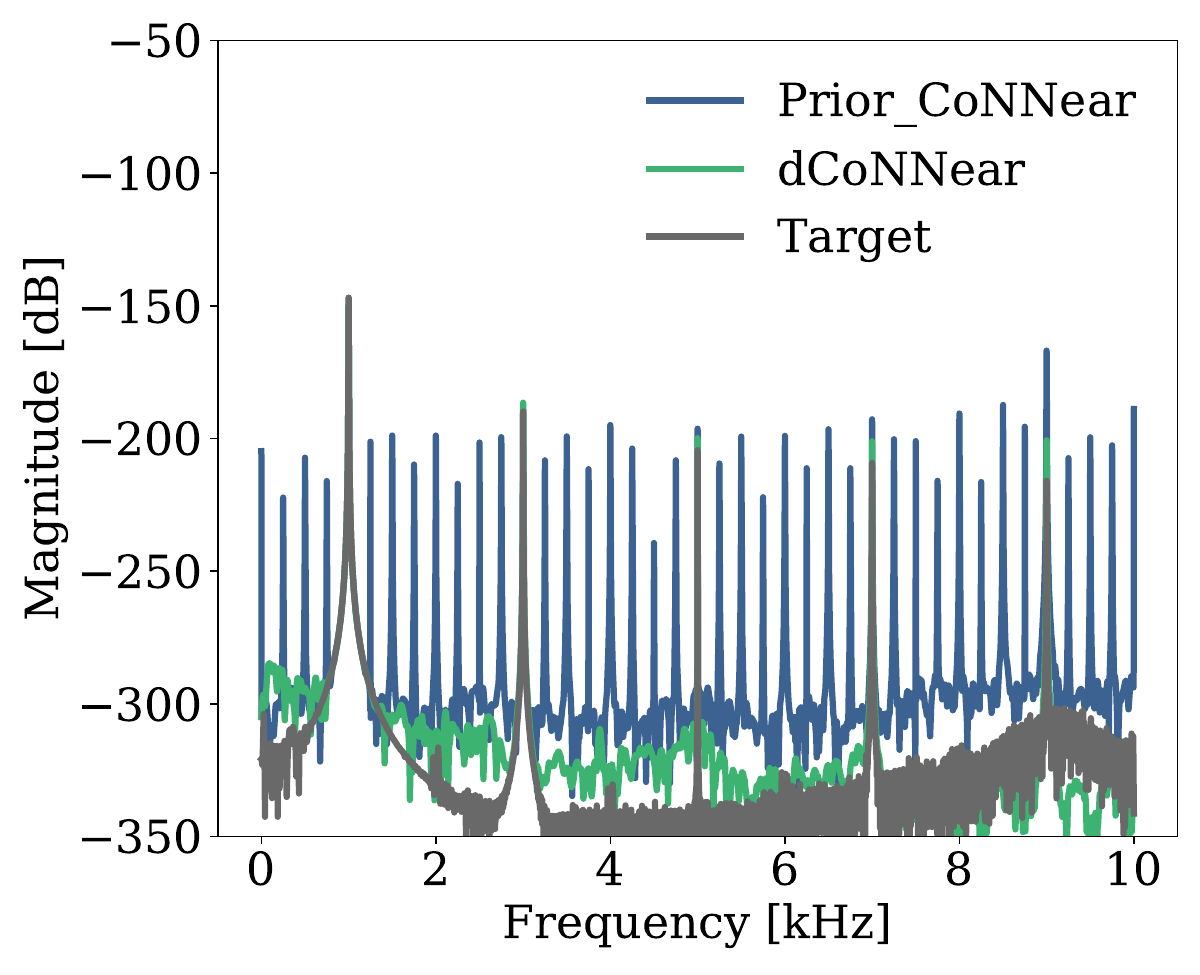}
\label{fig:ex3-a} }
\hspace{8pt}   
\subfloat[IHC outputs]{
\includegraphics[width=0.25\textwidth]{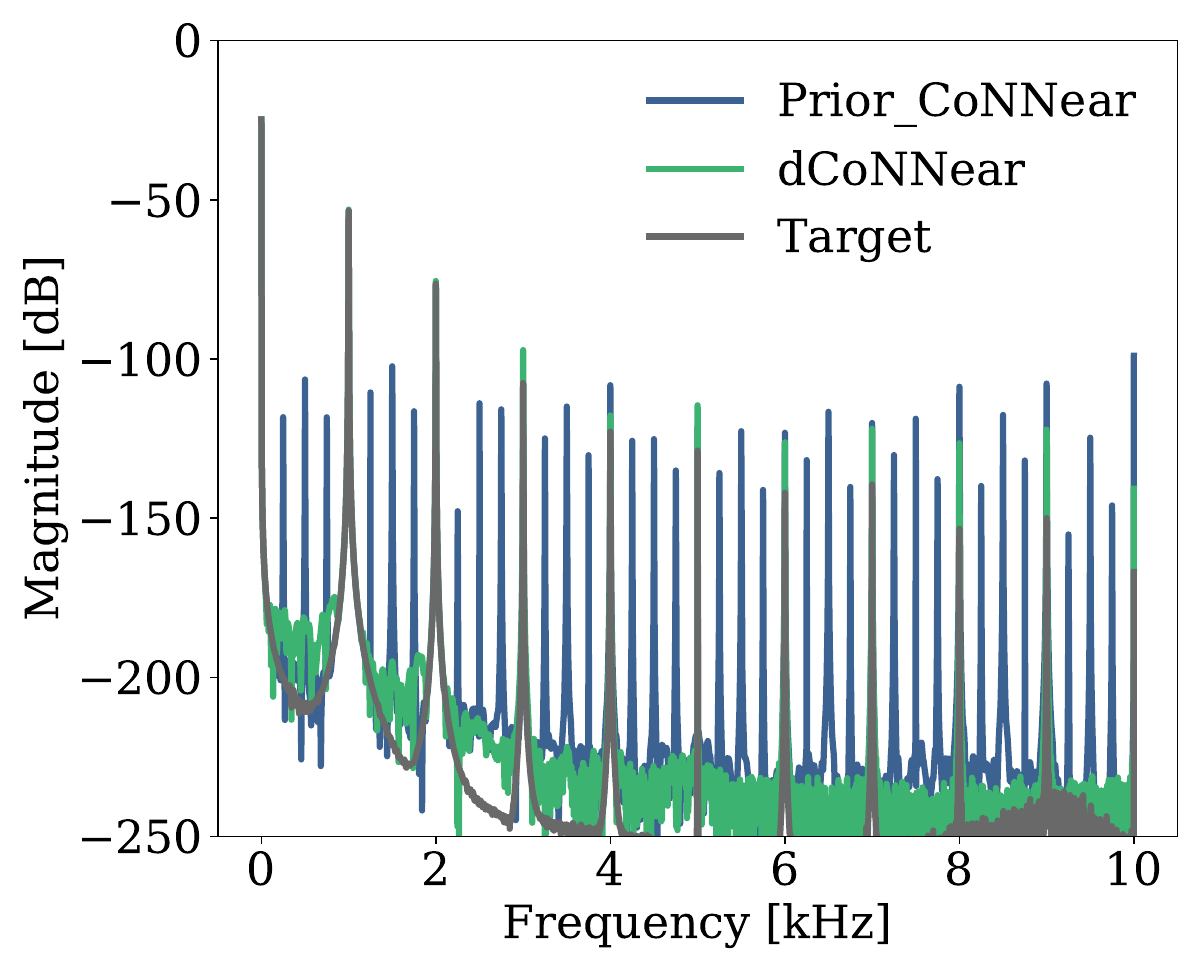}
\label{fig:ex3-b}}
\vspace{-10pt}

\subfloat[ANF outputs]{
\includegraphics[width=0.25\textwidth]{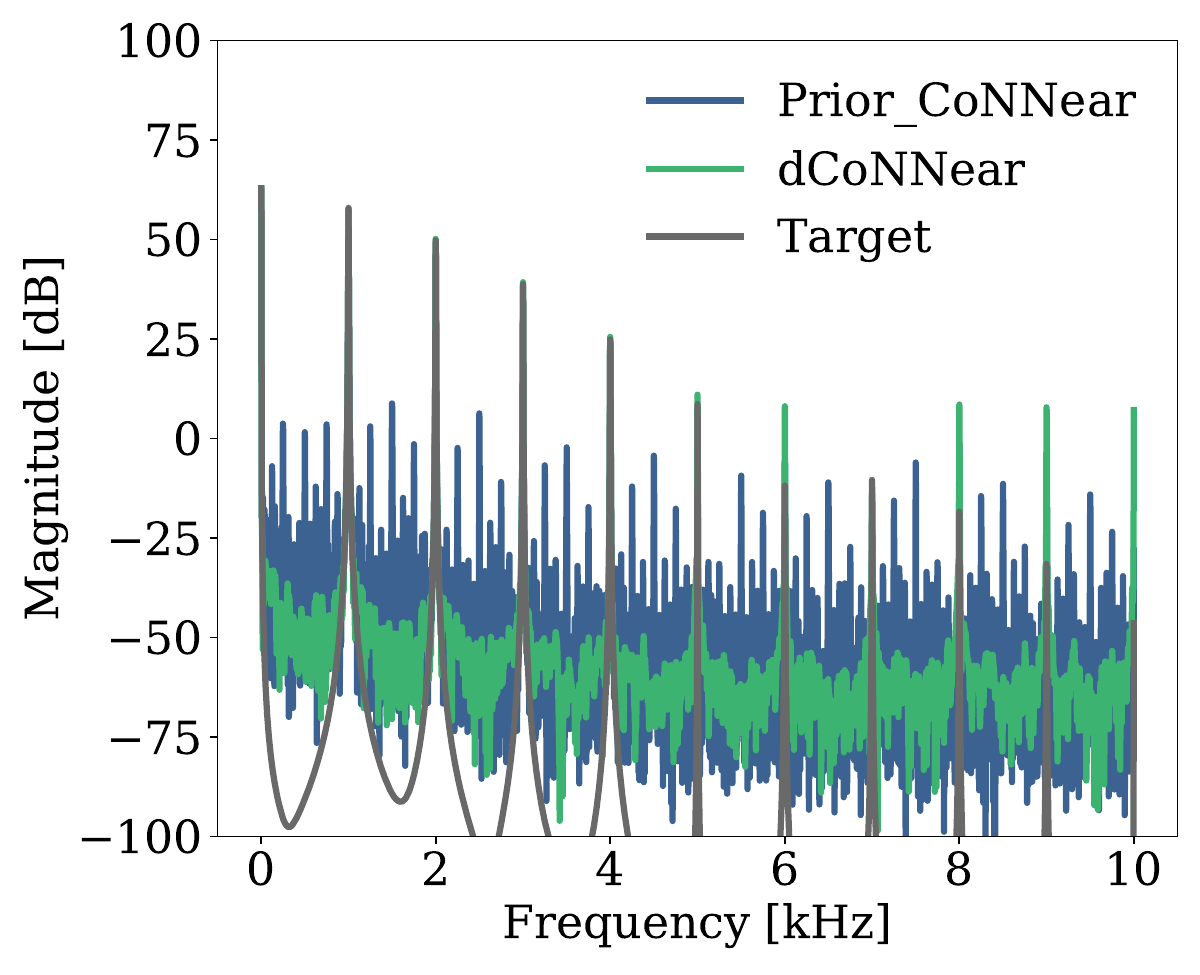}
\label{fig:ex3-c}}
\hspace{8pt} 
\subfloat[HA outputs]{
\includegraphics[width=0.25\textwidth]{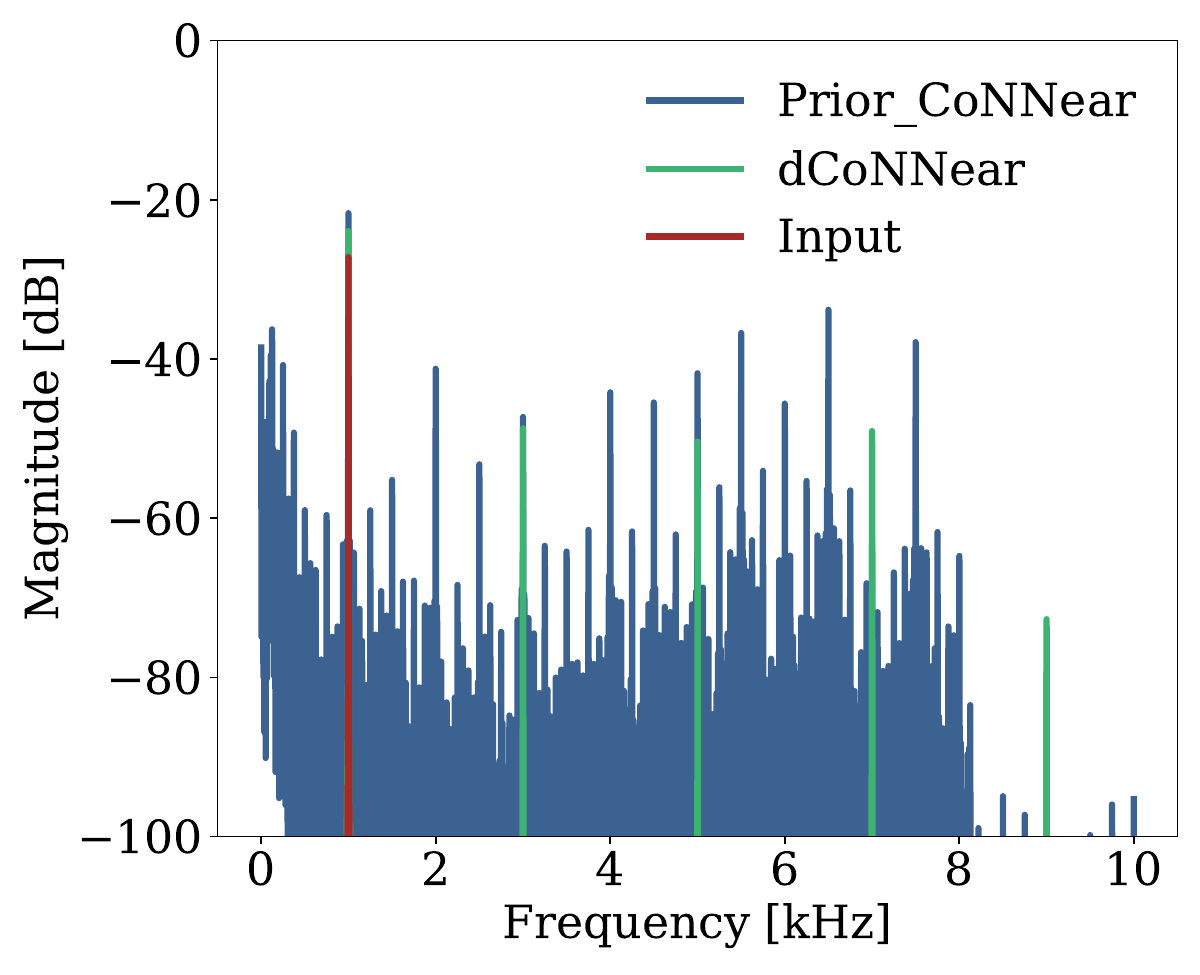}
\label{fig:ex3-d}}

\caption[]{The 1-kHz tone response at various auditory processing stages compared to the target models for normal hearing (a-c). (d) The 1-kHz tone responses of HA models trained using either the prior CoNNEar or dCoNNear architecture. The target refers to simulations with non-DNN based auditory models that formed the basis for the DNN-based CoNNEar model elements.}%
\label{fig:dCoNNear_artifacts_closedLoop}%
\vspace{-10pt}
\end{figure*}

\section{Results}
In this section, we evaluated the proposed artifact-free dCoNNear-based closed-loop framework, designed for individualized DNN-based hearing-aid (HA) models. The evaluation aimed to determine whether the artifacts associated with the prior CoNNear architecture were successfully eliminated in the dCoNNear-based auditory elements. We also compared the sound quality of the HA models trained between the prior CoNNear and new dCoNNear frameworks. Additionally, we assessed the biophysical properties of the auditory elements using stimuli that were not part of the training material. Finally, we compared the hearing-aid restoration performance of the models developed using the prior CoNNear-based framework with those developed using the dCoNNear framework.

\begin{table}[tb!]
\centering
\caption{Comparing the THD across different models for 1-kHz pure tone response}
\begin{tabular}{ccccc}
\hline
              & Cochlear & IHC    & ANF   & HA     \\ \hline
Analytical model        & -54.65   & -22.71 & -8.05 & -    \\
Prior coNNear & -29.66   & -20.47 & -4.64 & -28.06 \\
dCoNNear      & -32.92   & -22.01 & -8.16 & -39.78 \\ \hline
\label{tab:THD_sine1k}
\vspace{-10pt}
\end{tabular}
\end{table}

\begin{figure*}[tb!]
\centering
\includegraphics[width=0.9\textwidth,height=0.9\textheight,keepaspectratio]{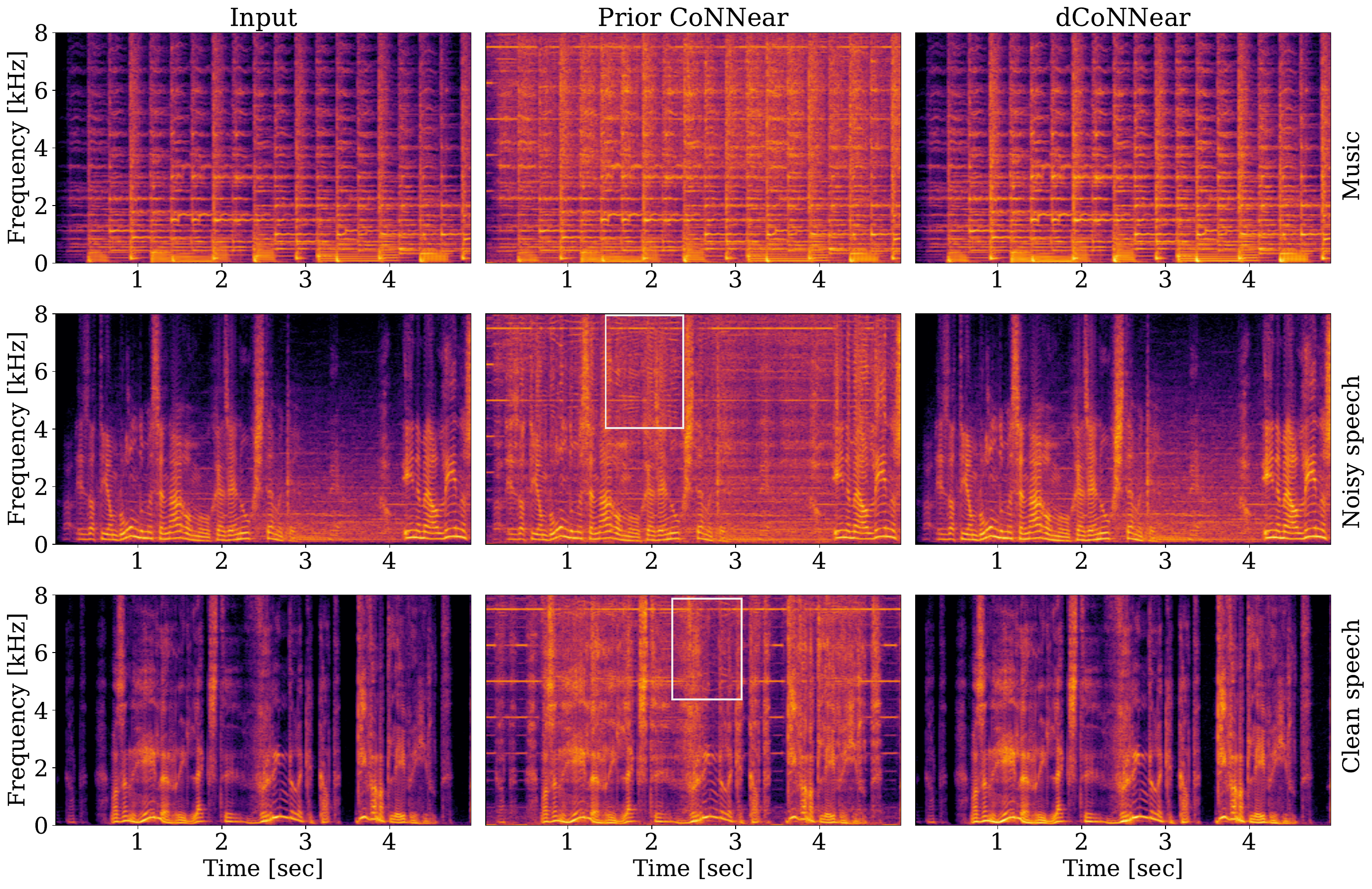}
\caption{Comparison of spectrograms processed by different HA models. The stimuli, from top to bottom, include music, clean speech, and noisy speech at 70 dB SPL.}
\label{sound_quality}
\vspace{-5pt}
\end{figure*}

\begin{figure}[tb!]
\centering
\includegraphics[width=0.4\textwidth,height=0.41\textheight,keepaspectratio]{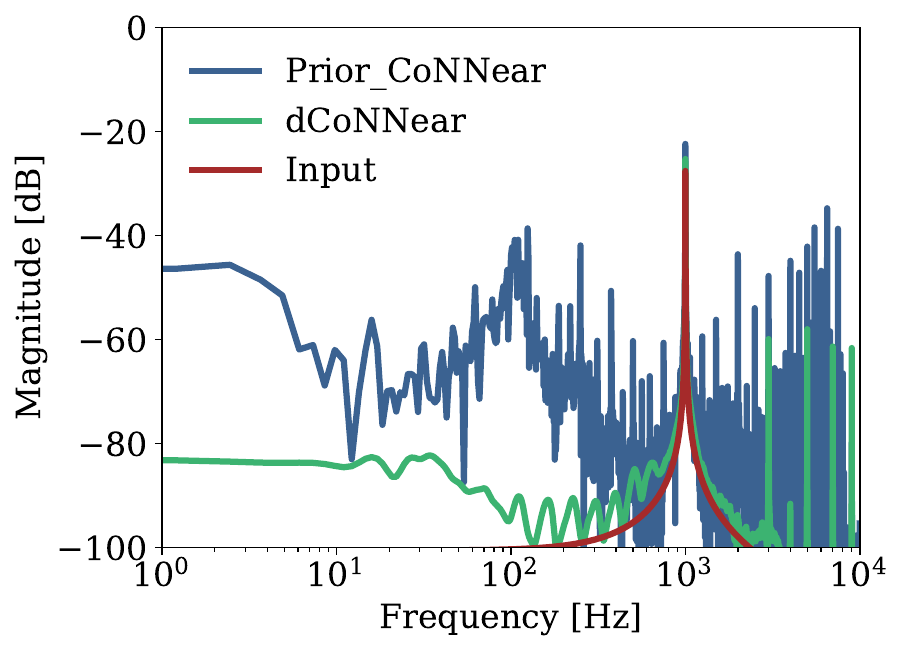}
\caption{Comparison of the frequency responses of a 1-kHz pure tone by different models.}
\label{fig:downsamp_aliasing}
\vspace{-10pt}
\end{figure}

\subsection{Artifacts and sound quality}
Artifacts were examined using total harmonic distortion (THD) at each stage of auditory processing and the resulting HA models in both the prior CoNNear and dCoNNear models. The sound quality of the HA models was evaluated using the non-intrusive metric SRMR on clean speech, noisy speech, and music samples.

Fig. \ref{fig:dCoNNear_artifacts_closedLoop} illustrates the 1-kHz tone responses at different auditory processing stages (panel a-c), as well as the HA models trained with diverse frameworks (panel d). The artifacts associated with the outputs of the dCoNNear models were markedly reduced compared to prior CoNNears, as observed by the difference between green and blue tonal lines in the magnitude spectrum. 
Table \ref{tab:THD_sine1k} shows the total harmonic distortions (THD) of the 1 kHz tone responses across different models. The THDs of the dCoNNear models decreased by 2 to 3.5 dB at each auditory processing stage compared to the prior CoNNear-based models, and the THD of the resulting HA outputs was reduced by 11.71 dB. This indicates that dCoNNear-based models effectively minimize the artifacts introduced by the prior CoNNear models, which enhances the resulting audio quality.

Table \ref{tab:quality_TIMIT} and \ref{tab:quality_LibrisTTS} compare the sound quality between prior CoNNear and dCoNNear for clean speech, noisy speech, and music, using two different corpora: TIMIT and LibriTTS.
Across both datasets, dCoNNear consistently outperformed prior CoNNear in terms of SRMR, DNSMOS, and PESQ. This consistent improvement in both clean and noisy conditions indicates that the dCoNNear-based model enhances the overall sound quality of the resulting speech signals. 
For music, however, the trend deviated slightly. While dCoNNear still outperformed prior CoNNear in DNSMOS and PESQ, its SRMR scores were marginally lower. 
This is likely due to the nature of music, which involves harmonics and overtones that are crucial for its perception. SRMR, however, primarily focuses on human vocal frequencies and modulations, and may not fully capture the complexities of music.

Fig. \ref{sound_quality} presents spectrogram examples across different audio types for both model architectures. The prior CoNNear models exhibited prominent tonal artifacts, visible as horizontal lines over time, which were absent in the original inputs. These artifacts were especially pronounced in clean speech and could be attributed to two primary structural issues: (1) tonal artifacts introduced by transposed convolution layers and (2) spectral replication, commonly known as imaging artifacts, caused by naive upsampling operations.
In the spectrograms of speech processed by the prior CoNNear-based model, equally spaced mirror-image bands appeared in the high-frequency regions (highlighted by the white rectangular box). These image artifacts resulted from spectral replication during upsampling. In contrast, the horizontal lines and replicated high-frequency bands were minimized in the dCoNNear outputs, indicating that the architecture effectively suppressed these types of artifacts.
\\
Fig. \ref{fig:downsamp_aliasing} shows the frequency responses for a 1 kHz pure tone input processed by both models. The prior CoNNear output (blue) exhibited dense clusters of spurious low-frequency peaks below 500 Hz, which were absent in both the input and the dCoNNear output. These peaks reflected aliasing artifacts caused by repeated downsampling without adequate anti-aliasing filtering. Specifically, the prior CoNNear-based HA model includes eight downsampling layers with stride 2, reducing the effective Nyquist frequency at the deepest layer to approximately 39 Hz. As a result, the original 1 kHz tone was repeatedly folded into lower frequency bands at each stage, accumulating into artificial low-frequency components. These artifacts could perceptually manifest as low-frequency buzzing or modulation noise. In contrast, dCoNNear avoids the aliasing artifacts due to the problematic downsampling, yielding a cleaner and perceptually more natural output.

The resulting audio samples of music, clean speech, and noisy speech are provided online \footnote{\url{https://github.com/chuan997/Trans24_audio_samples}}.

\begin{table}[tb!]
\centering
\caption{Comparison of sound quality across different HA models where the clean speech corpus is from TIMIT dataset.}
\begin{tabular}{ccccc}
\hline
Aduio type                    & Model         & SRMR & DNSMOS & PESQ \\ \hline
\multirow{3}{*}{Clean speech} & Unprocessed   & 7.39 & 3.33   & 4.64 \\
                              & prior CoNNear & 6.13 & 2.22   & 1.82 \\
                              & dCoNNear      & 6.77 & 3.26   & 4.31 \\ \hline
\multirow{3}{*}{Noisy speech} & Unprocessed   & 6.45 & 2.64   & 2.68 \\
                              & prior CoNNear & 5.17 & 2.06   & 2.34 \\
                              & dCoNNear      & 6.24 & 2.52   & 2.62 \\ \hline
\multirow{3}{*}{Music}        & Unprocessed   & 1.93 & 1.09   & -    \\
                              & prior CoNNear & 1.91 & 1.08   & 1.75 \\
                              & dCoNNear      & 1.75 & 1.07   & 2.25 \\ \hline
\end{tabular}
\label{tab:quality_TIMIT}
\end{table}

\begin{table}[tb!]
\centering
\caption{Comparison of sound quality across different HA models where the clean speech corpus is from LibriTTS dataset.}

\begin{tabular}{ccccc}
\hline
Aduio type                    & Model         & SRMR & DNSMOS & PESQ \\ \hline
\multirow{3}{*}{Clean speech} & Unprocessed   & 7.84 & 3.49   & 4.64 \\
                              & prior CoNNear & 6.48 & 2.24   & 1.97 \\
                              & dCoNNear      & 7.01 & 3.18   & 4.22 \\ \hline
\multirow{3}{*}{Noisy speech} & Unprocessed   & 6.71 & 2.57   & 2.73 \\
                              & prior CoNNear & 5.46 & 2.11   & 2.24 \\
                              & dCoNNear      & 6.41 & 2.49   & 2.61 \\ \hline
\end{tabular}
\label{tab:quality_LibrisTTS}
\end{table}

\subsection{Biophysical properties of auditory modules}
To ensure that changing the CoNNear architecture did not compromise the performance,
we employed auditory neuroscience techniques to quantify key properties of the dCoNNear-based auditory elements, comparing them against the reference analytical model and prior CoNNear outputs.

\begin{figure*}[tb!]
\centering
\includegraphics[width=0.9\textwidth,height=0.9\textheight,keepaspectratio]{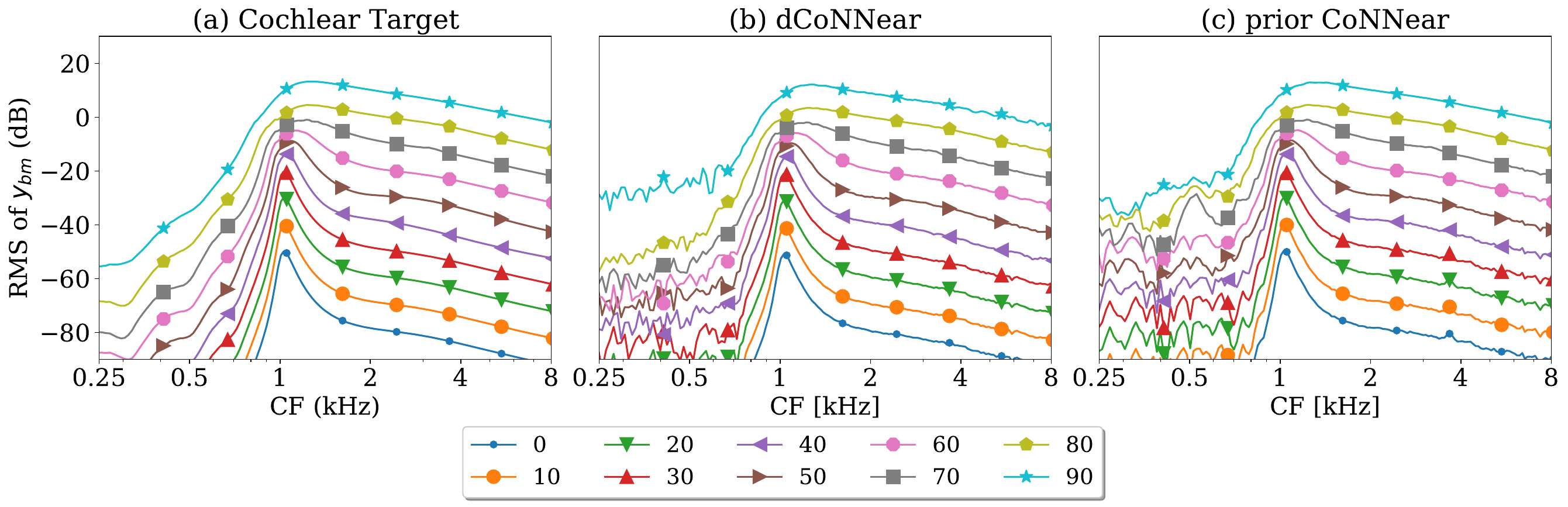}
\caption{Comparison of simulated RMS levels of basilar membrane (BM) displacement across center frequencies (CF) for pure-tone stimuli between 0 and 90 dB SPL with a step of 10 dB: (a) Target model, (b) dCoNNear model, (c) prior CoNNear model. }
\label{cochlear_excitation}
\vspace{-15pt}
\end{figure*}

\begin{figure}[tb!]
\centering
\includegraphics[width=0.35\textwidth]{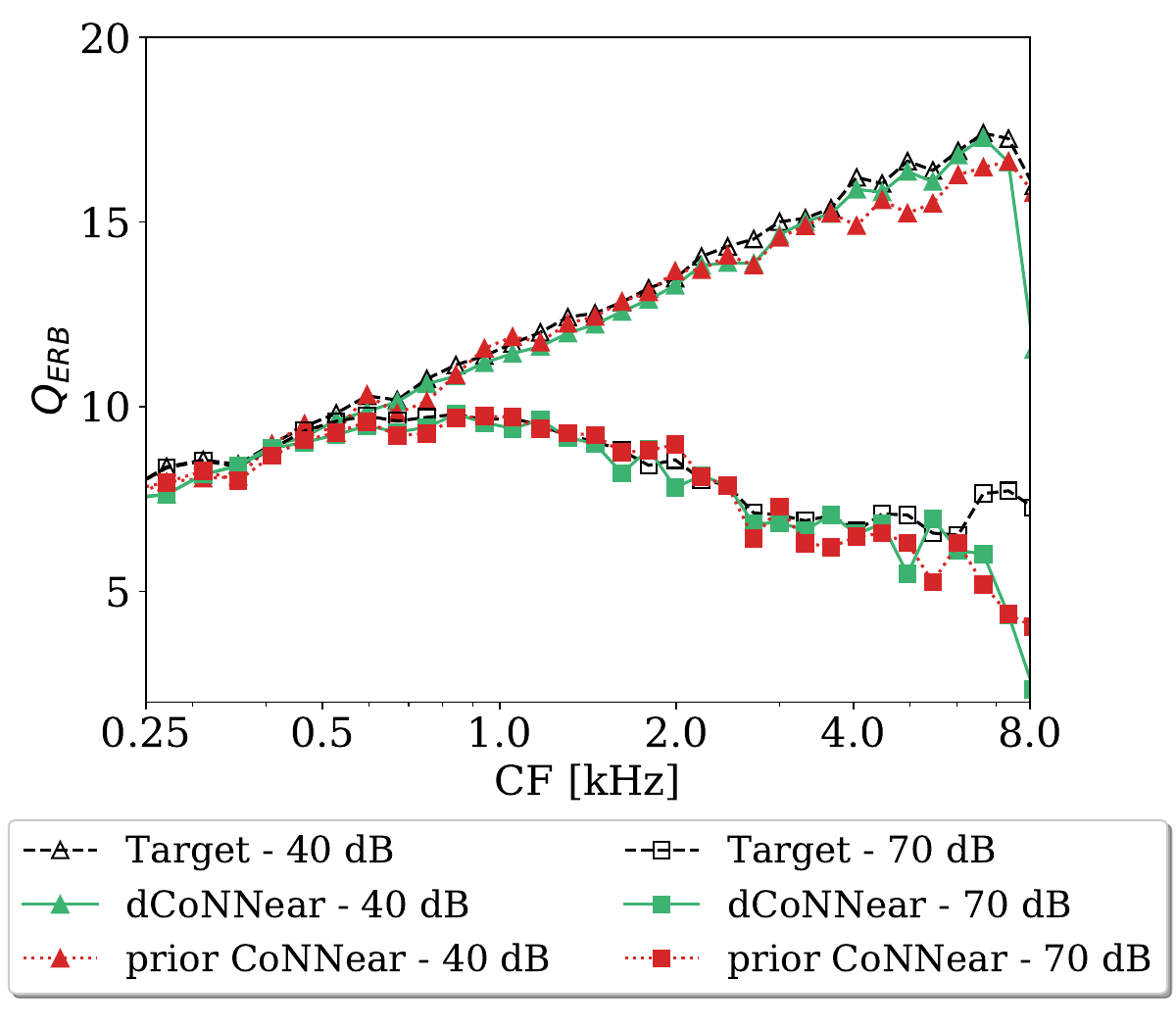}
\caption{Comparison of cochlear filter tuning (Q\textsubscript{ERB}) across various center frequencies (CFs) using 100 $\mu$s clicks at 40 and 70 dB SPL among the dCoNNear model, the prior CoNNear model, and the target model. }
\label{cochlear_QERB}
\vspace{-10pt}
\end{figure}

Figures \ref{cochlear_excitation} and \ref{cochlear_QERB} compare the basilar membrane (BM) excitation patterns and Q\textsubscript{ERB} values of the proposed dCoNNear model with the reference TL model and prior CoNNear\textsubscript{cochlear}. As shown in Fig. \ref{cochlear_excitation}, dCoNNear exhibited performance comparable to both the prior CoNNear model and the TL model, suggesting that dCoNNear accurately captures the shape and compression properties of pure-tone cochlear excitation patterns. Fig. \ref{cochlear_QERB} illustrates a close match between dCoNNear, the reference TL model, and CoNNear\textsubscript{cochlear}, indicating that dCoNNear reliably replicates the frequency-dependent characteristics of cochlear responses.

\begin{figure*}[tb!]
\centering
\includegraphics[width=0.85\textwidth,height=0.9\textheight,keepaspectratio]{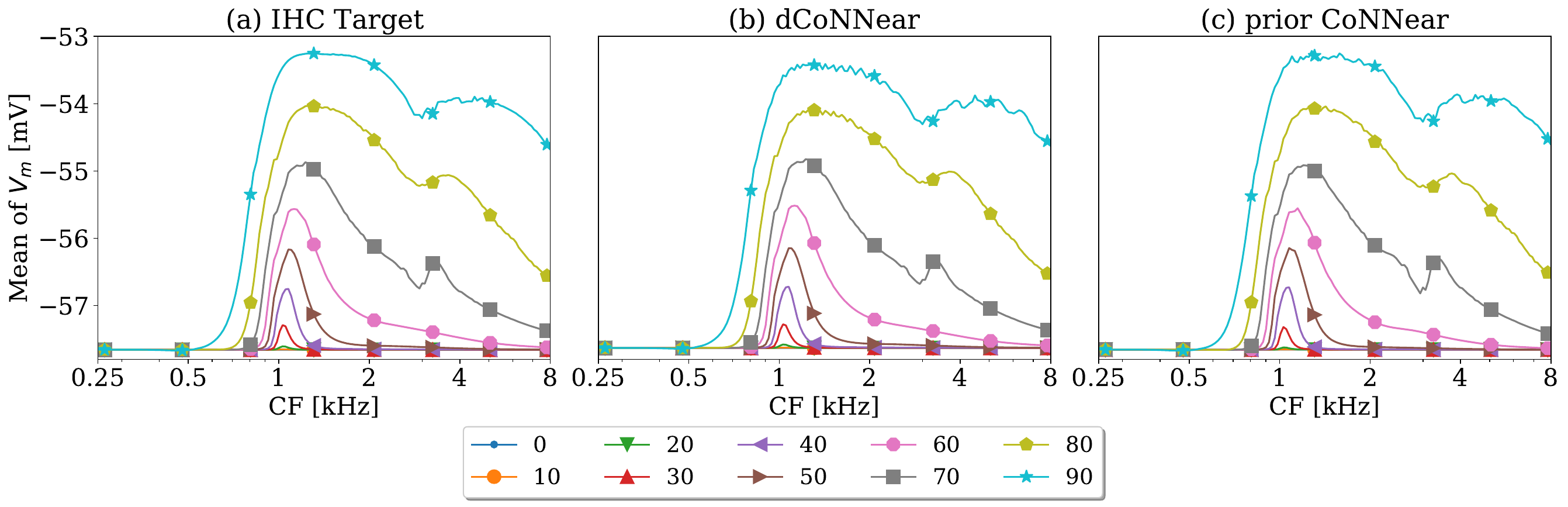}
\caption{
Comparison of average IHC receptor potentials  V\textsubscript{m} across CFs for 1-kHz tone stimuli 0 to 90 dB SPL with a step of 10 dB: (a) target model, (b) dCoNNear model, and (c) prior CoNNear model.}
\label{ihc_ex_pattern}
\end{figure*}

\begin{figure}[tb!]
\centering
\includegraphics[width=0.28\textwidth]{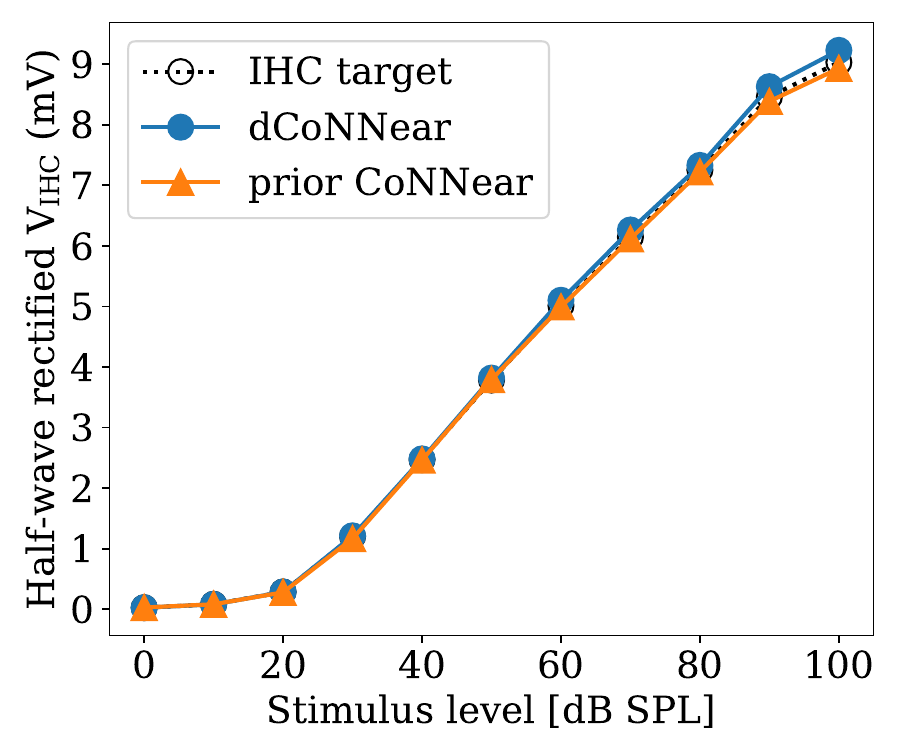}
\caption{Comparison of simulated dCoNNear\textsubscript{IHC} half-wave rectified receptor potential in response to a 4-kHz pure tone across varying sound levels: (a) Target model, (b) dCoNNear model, (c) prior CoNNear model.}
\label{ihc_half_wave}
\vspace{-10pt}
\end{figure}

Figs. \ref{ihc_ex_pattern} and \ref{ihc_half_wave} present the IHC excitation patterns and the half-wave rectified IHC receptor potential, compared to the reference IHC model and prior CoNNear\textsubscript{IHC} predictions. The excitation patterns from the dCoNNear\textsubscript{IHC} model closely align with those of both the reference IHC model and prior CoNNear\textsubscript{IHC} (Fig. \ref{ihc_ex_pattern}). Fig. \ref{ihc_half_wave} shows the RMS of the half-wave rectified IHC receptor potential, $V\textsubscript{IHC}$, in response to a 4-kHz pure tone across input levels ranging from 0 to 100 dB SPL. The blue curve, representing the dCoNNear\textsubscript{IHC} model, follows the trends of the black and orange curves corresponding to the reference and prior models, respectively. This consistency indicates a nearly linear relationship with SPL up to 90 dB, followed by compressive growth beyond 90 dB SPL. These results suggest that the dCoNNear\textsubscript{IHC} model accurately captures the potential-level growth characteristics.

\begin{figure}[tb!]
\centering
\includegraphics[width=0.35\textwidth]{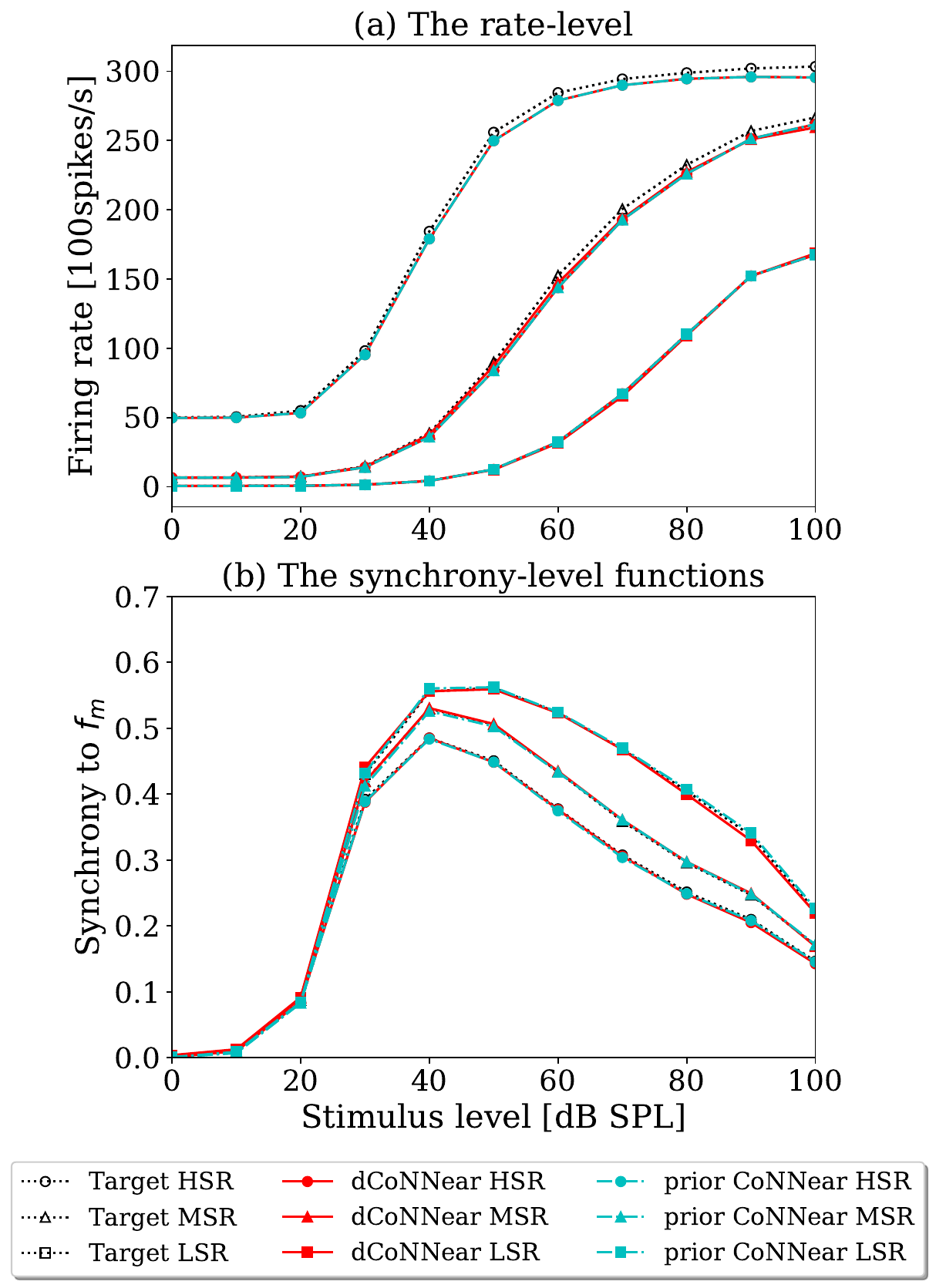}
\caption{Comparison of evaluation metrics for different emulated ANF models: Target, dCoNNear, and prior CoNNear. (a) ANF rate-level curves simulated for the HSR, MSR, and LSR ANF models at a CF of 4 kHz. (b) ANF synchrony-level functions for the HSR, MSR, and LSR ANF models at a CF of 4 kHz in response to a 4 kHz tone modulated by a 100 Hz tone.}
\label{ANF_combine}
\vspace{-10pt}
\end{figure}

Fig. \ref{ANF_combine}(a) and (b) illustrate the ANF rate-level curves and synchrony levels, respectively, compared to the reference ANF model and prior CoNNear\textsubscript{ANF} predictions. In Fig. \ref{ANF_combine}(a), the dCoNNear\textsubscript{ANF} model showed consistent growth with the reference ANF model and prior CoNNear\textsubscript{ANF} for HSR, MSR, and LSR types. This indicates that the dCoNNear\textsubscript{ANF} model captures the level-dependent properties of different ANF types. Fig. \ref{ANF_combine}(b) demonstrates that the dCoNNear\textsubscript{ANF} model accurately represents the synchrony-level curves of the reference ANF model and prior CoNNear\textsubscript{ANF} for all ANF types.

In conclusion, the dCoNNear-based auditory models faithfully replicate the auditory properties of the reference analytical models, ensuring accurate and reliable performance in auditory signal processing tasks even with the new dCoNNear architecture.

\subsection{Evaluation of restoration performance}
\begin{figure*}[tbp]
\centering
\includegraphics[width=0.75\textwidth]{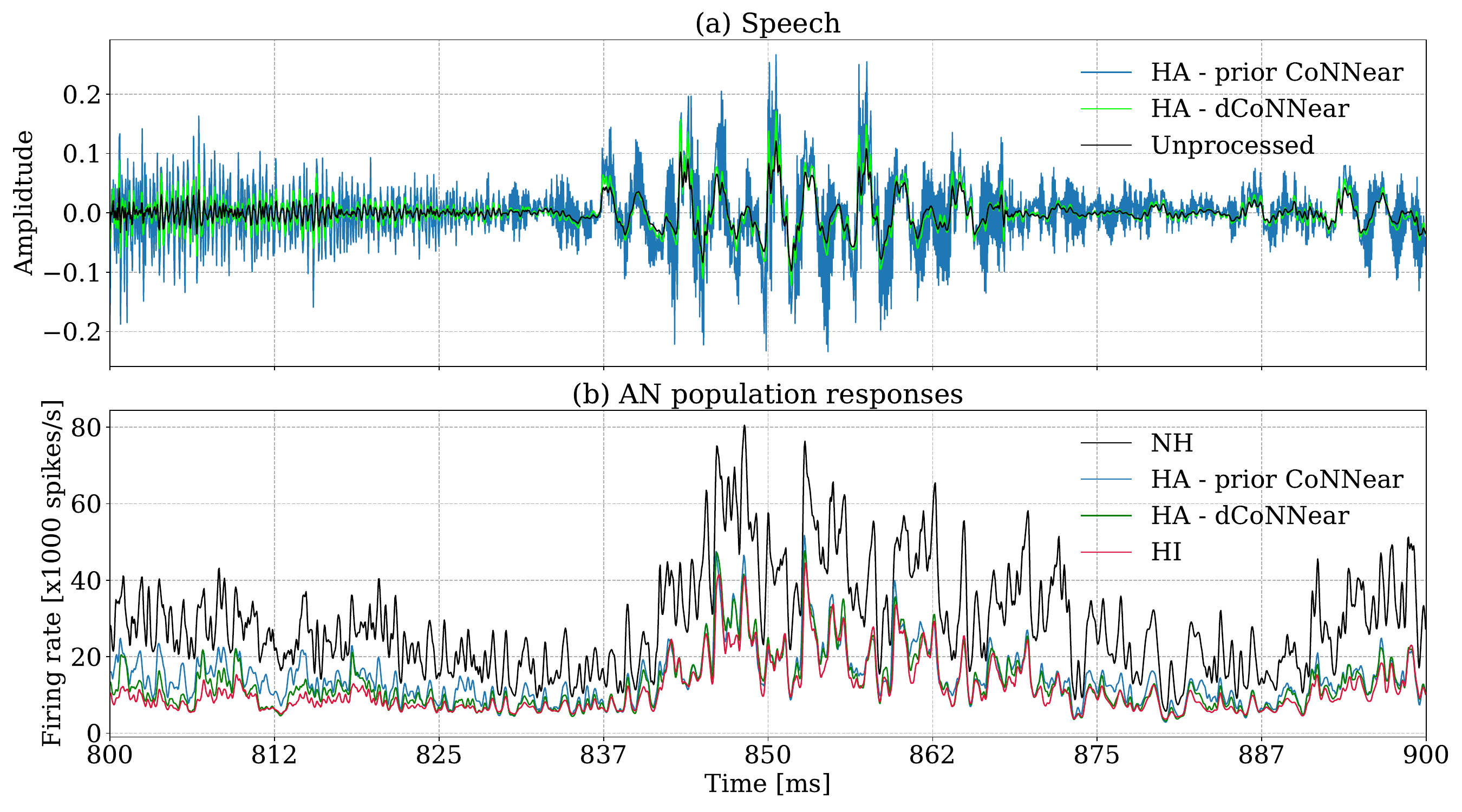}
\caption{Simulated auditory nerve (AN) population responses for hearing-aid (HA) models trained using different frameworks at 70 dB SPL. (a) A clean speech segment is shown before and after processing with models trained using prior CoNNear (blue) and dCoNNear (green) systems. (b) The hearing-impaired (HI) AN population responses were enhanced after processing the input stimulus with different HA models. Ideal compensation would align the HI AN population responses with the NH AN population responses.}
\label{fig: compensation}
\vspace{-10pt}
\end{figure*}

The restoration performance of trained HA models based on different frameworks was evaluated and are presented in Table \ref{tab:NRMSE} and Fig. \ref{fig: compensation}.

Table \ref{tab:NRMSE} presents the average normalized root-mean-square errors (NRMSEs) between the normal NH and HI AN population responses for utterances from the TIMIT and LibriTTS corpora. These values were calculated for  unprocessed and processed signals using hearing-aid (HA) models trained with either the prior CoNNear or the proposed dCoNNear frameworks. To ensure a fair comparison, both HA models were trained using the same loss function and optimization settings. 
In both corpora, the results followed consistent trends. Both the prior CoNNear and dCoNNear-based HA models effectively reduced the NRMSE across input levels ranging from 40 to 70 dB SPL, with the largest improvements observed at lower input levels (40–50 dB SPL), where HA processing plays a more significant role.
However, the prior CoNNear exhibited marginally better performance than the dCoNNear. This may be attributed to the tonal artifacts generated by the HA models trained from the prior CoNNear-based system. As noted in \cite{drakopoulos2022model}, temporal peaks or spurious high-frequency tones can result in increased AN responses, potentially narrowing the gap between impaired and reference NH responses in a way that lowers NRMSE but does not reflect true restoration.

To better understand how the processed audio affects the AN response features, Fig. \ref{fig: compensation}  illustrates the time-domain speech segment along with the corresponding AN population responses before and after processing. A speech segment was processed by the two HA models and used as input to the HI auditory systems to simulate their AN population responses. The HA model trained using the prior CoNNear-based system introduced additional high-frequency fluctuations to the stimulus, enhancing the AN population responses but not fully restoring them. Conversely, the dCoNNear-based HA model introduced significantly fewer high-frequency fluctuations, resulting in a comparatively smaller increase in AN population responses. The additional high-frequency fluctuations stemmed from tonal artifacts introduced by the prior CoNNear, which increased AN population responses, but degraded the sound quality.
This suggests that the tonal artifacts introduced by prior CoNNear contribute to restoring HI AN population responses, potentially enhancing hearing loss compensation. However, they compromised the sound quality of the resulting HA models. 
In conclusion, the dCoNNear-based models demonstrated comparable restoration performance while minimizing the artifacts introduced by prior CoNNear-based models, thereby achieving better overall sound quality.

\subsection{Real-Time Inference Evaluation}
To assess the real-time performance of the proposed audio processor, we measured the inference latency on an Intel i7-1265U CPU and an NVIDIA A30 GPU. A total of 100 speech samples from the test set were processed, each consisting of 512 samples (equivalent to 25.6 ms at a 20 kHz sampling rate). The dCoNNear-based HA model achieved average processing times of 16.42 ms on the CPU and 3.81 ms on the GPU, corresponding to real-time factors (RTFs) of 0.64 and 0.15, respectively, indicating real-time capability on both platforms. For comparison, the previous CoNNear-based HA model from \cite{drakopoulos2023neural} processed the same input frames in 3.98 ms on the CPU and 1.57 ms on the GPU, yielding RTFs of 0.15 and 0.06, respectively.
While the dCoNNear-based model introduces additional computational complexity compared to the original CoNNear model, it still meets real-time constraints, even on a standard CPU. 

\begin{table}[tb!]
\centering
\caption{Average NRMSE (\%) across different HA models for input levels between 40 and 70 dB SPL, evaluated on the TIMIT and LibriTTS corpora.}
\begin{tabular}{cccccc}
\hline
Dataset                    & Model         & 40    & 50    & 60    & 70    \\ \hline
\multirow{3}{*}{TIMIT}     & Unprocessed   & 29.08 & 20.13 & 15.86 & 14.25 \\
                           & prior CoNNear & 20.13 & 15.41 & 14.67 & 13.12 \\
                           & dCoNNear      & 20.21 & 15.54 & 15.28 & 13.65 \\ \hline
\multirow{3}{*}{LibrisTTS} & Unprocessed   & 30.04 & 19.14 & 15.08 & 14.09 \\
                           & prior CoNNear & 20.87 & 15.73 & 14.29 & 13.24 \\
                           & dCoNNear      & 21.16 & 15.81 & 14.35 & 13.26 \\ \hline
\label{tab:NRMSE}
\end{tabular}
\end{table}

\begin{figure}[tb!]
\centering
\includegraphics[width=0.48\textwidth]{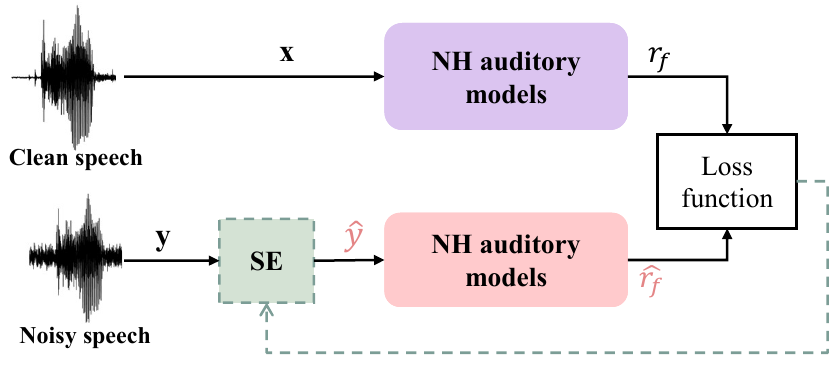}
\caption{The framework of training speech enhancement models. The auditory modules are based on prior CoNNear and dCoNNear.} 
\label{fig:NR_framework}
\vspace{-10pt}
\end{figure}

\section{Speech Enhancement Application}
Beyond its primary validation in hearing-aid design, the dCoNNear framework's inherent ability to generate artifact-free audio extends its utility to various wave-to-wave audio processing tasks. This section, therefore, presents a proof-of-concept experiment to demonstrate dCoNNear's efficacy in a closed-loop speech enhancement (SE) task, illustrating its broader applicability.
\subsection{Framework}
The training framework for the speech enhancement (SE) model is illustrated in Fig. \ref{fig:NR_framework}. Both processing pathways include auditory modules that simulate normal-hearing auditory profiles. In the reference pathway, clean speech is fed directly into the auditory system. In the parallel pathway, the speech enhancement model first processes a noisy input, and its output is then passed through the same auditory modules. Both the auditory modules and the SE model follow the same architectural design as those used in the hearing-aid application described earlier.
\subsection{Training}
We used utterances from the INTERSPEECH 2021 DNS Challenge dataset \cite{reddy2021interspeech} for training and utterances from LibriTTS \cite{zen2019libritts} for testing. For noise data, Audioset \cite{gemmeke2017audio} was used during training, while Freesound \cite{fonseca2017freesound} served as the test noise source. Noisy speech samples were generated by randomly mixing clean utterances with noise segments at signal-to-noise ratios (SNRs) ranging from –5 dB to 5 dB in 1 dB increments. This resulted in 100 hours of training data and 5 hours of validation data. A separate 1-hour test set was created at each of the three SNR levels: –5 dB, 0 dB, and 5 dB. All samples were resampled to 20 kHz to match the frequency range of the auditory models and normalized to 70 dB SPL to ensure consistency with the closed-loop system.
\\
The SE model was trained to minimize the mean absolute error (MAE) between the outputs of the two auditory pathways. During training, the parameters of the auditory modules were kept frozen. For comparison, we also trained a prior CoNNear-based SE model using the same training procedure and dataset. This allowed for a direct evaluation of the benefits introduced by the dCoNNear architecture in a closed-loop speech enhancement task.
\\
\subsection{Results}
\begin{figure}[tb!]
\centering
\includegraphics[width=0.45\textwidth]{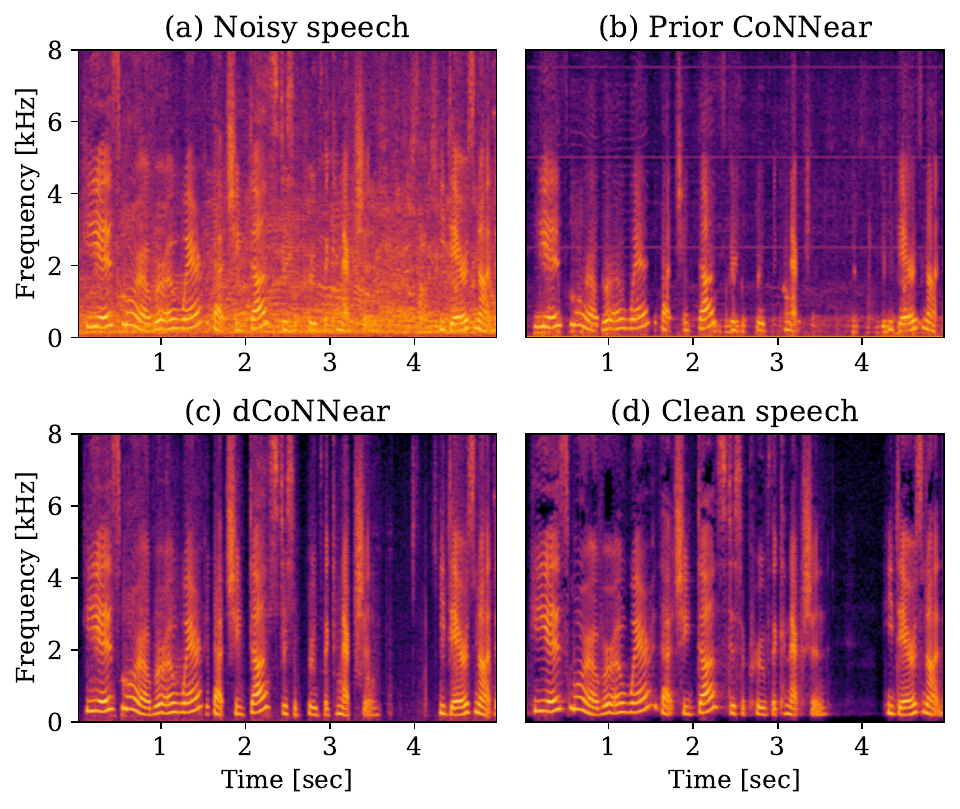}
\caption{Spectrograms of (a) noisy speech at 0 dB SNR with babble noise, (b) enhancement by prior CoNNear-based model, (c) enhancement by dCoNNear-based model, and (d) clean speech.}
\label{fig:quality_NR}
\vspace{-10pt}
\end{figure}

We evaluated the sound quality of the prior CoNNear- and dCoNNear-based speech enhancement (SE) models using three objective metrics: DNSMOS (overall) \cite{reddy2021dnsmos}, PESQ \cite{rix2001perceptual}, and ESTOI \cite{jensen2016algorithm}. Table \ref{tab:quality_NR} presents the performance comparison across input SNRs of –5 dB, 0 dB, and 5 dB.
Across all SNR conditions, the dCoNNear-based SE model consistently achieved higher scores in DNSMOS, PESQ, and ESTOI, indicating improved perceptual quality and intelligibility over the prior CoNNear-based model.
\\
To further analyze the presence of artifacts, Fig.~\ref{fig:quality_NR} displays spectrograms of a representative speech example corrupted by babble noise and subsequently enhanced by the two SE models. The output of the prior CoNNear-based model exhibited distinct tonal artifacts, visible as horizontal lines in the spectrogram, which were not present in the original noisy input. In contrast, the spectrogram from the dCoNNear-based model showed a marked reduction in these artifacts. This qualitative evidence supported the objective results and demonstrated that dCoNNear effectively reduces structural artifacts in the closed-loop speech enhancement task. The resulting audio samples are provided online \footnote{\url{https://github.com/chuan997/Trans24_audio_samples}}.
\\
The results confirm the effectiveness of dCoNNear as a robust and generalizable solution for artifact-free closed-loop audio processing, extending its applicability beyond hearing-aid processing to speech enhancement tasks and potentially other wave-to-wave audio domains.

\begin{table}[tb!]
\centering
\caption{Comparison of different models in terms of various objective metrics. Higher values indicate better performance and bold fonts highlight the best performance.}
\begin{tabular}{ccccc}
\hline
\multirow{2}{*}{Metrics} & \multirow{2}{*}{Models} & \multicolumn{3}{c}{SNRs (dB)}                 \\ \cline{3-5} 
                         &                         & -5            & 0             & 5             \\ \hline
\multirow{3}{*}{DNSMOS}  & Unprocessed             & 1.08          & 1.29          & 1.5           \\
                         & prior CoNNear           & 1.57          & 2.32          & 2.54          \\
                         & dCoNNear                & \textbf{2.87} & \textbf{3.01} & \textbf{3.22} \\ \hline
\multirow{3}{*}{PESQ}    & Unprocessed             & 1.04          & 1.27          & 1.45          \\
                         & prior CoNNear           & 1.18          & 1.44          & 1.65          \\
                         & dCoNNear                & \textbf{1.3}  & \textbf{1.63} & \textbf{2.12} \\ \hline
\multirow{3}{*}{ESTOI}   & Unprocessed             & 0.35          & 0.48          & 0.6           \\
                         & prior CoNNear           & 0.39          & 0.59          & 0.66          \\
                         & dCoNNear                & \textbf{0.6}  & \textbf{0.74} & \textbf{0.82} \\ \hline
\end{tabular}
\label{tab:quality_NR}
\end{table}

\section{Discussion}
Current closed-loop frameworks suffer from non-ideal sampling processes, which can introduce undesired artifacts when placed in closed-loop frameworks and can degrade the audio quality of the resulting application. In this study, we systematically characterized the artifacts generated by different upsampling methods and examined those associated with transposed convolution within the context of closed-loop, autoencoder-based hearing-aid algorithms. These artifacts originated from the interaction between problematic upsampling operations and spectral replicas by the upsampling layers. The artifacts persisted in the closed-loop system after training, leading to tonal artifacts in the resulting audio.

We propose an artifact-free architecture, dCoNNear, designed to integrate into closed-loop systems, exemplified here for hearing-aid algorithms. In contrast to the existing autoencoder-based closed-loop system \cite{drakopoulos2023neural}, the dCoNNear-based architecture can significantly reduce the artifacts associated with prior CoNNear models, thereby improving the sound quality of the resulting audio. This advancement provides an artifact-free framework for audio applications, delivering high sound quality for future audio processing algorithms.

From the perspective of the human auditory system, dCoNNear's modifications significantly enhance fidelity by eliminating DNN architectural artifacts, such as tonal and aliasing distortions, which are inherently absent in natural human auditory processing.These artifacts can compromise the accuracy of bio-inspired models and distort perceptual outcomes. By removing them, dCoNNear enables a more faithful simulation of auditory function, resulting in artifact-free representations. This is particularly important for applications such as personalized hearing-aid algorithms, where the goal is to restore simulated auditory-nerve responses in hearing-impaired models to closely match those of normal-hearing references. Experimental results confirm that dCoNNear accurately models advanced biophysical stages without introducing spurious artifacts, achieving stronger alignment with the functional properties of the human auditory pathway.

While dCoNNear-based models successfully minimize artifacts associated with prior CoNNear, the SRMR scores of the processed samples decreased relative to the inputs, as shown in Table \ref{tab:quality_TIMIT} and \ref{tab:quality_LibrisTTS}. One potential reason is the compressive amplification characteristics of the trained HA model, where quieter sounds are amplified more than louder sounds. This process amplifies low-level noise, which subsequently degrades sound quality. Therefore, incorporating a noise reduction module should be considered in future works. Another contributing factor could be the training process, which aimed to minimize differences between NH and HI auditory processing but may introduce non-linear distortions that affect sound quality. This study underscores the need for future research focused on developing more optimal loss functions or post-processing filters. Such developments would enhance the perceptual benefits of audio processing systems. 


The use of the Speech-to-Reverberation Modulation Energy Ratio (SRMR) as an evaluation metric provided valuable insights into the sound quality of speech samples. However, SRMR falls short in evaluating other sounds types, such as music and environmental noises. This study suggests the need for broader objective metrics to assess a broader range of audio types.

While user-centric evaluation methods such as the Hearing in Noise Test (HINT) [37] and Multiple Stimuli with Hidden Reference and Anchor (MUSHRA) [38] offer valuable insights into perceptual experience, the present work focuses on addressing architectural artifacts in DNN-based audio models. We therefore adopt established objective metrics that correlate with perceived quality and intelligibility. Given the proof-of-concept nature of this work, integrating formal listening tests remains a valuable direction for future research, particularly as the system advances toward real-world deployment.


\section{Conclusion}
In this work, we proposed dCoNNear, a novel architecture specifically designed to suppress spurious artifacts—most notably tonal and aliasing artifacts introduced by non-ideal downsampling and upsampling layers. We demonstrated its effectiveness in closed-loop applications for both hearing-aid algorithms and speech enhancement. The dCoNNear architecture incorporates a series of stacked FIR-like memory blocks, eliminating the need for sampling operations and demonstrating strong capability in modeling long-term dependencies.
Our results showed that dCoNNear-based models significantly reduced the artifacts associated with prior autoencoder-based CoNNear systems and improved perceptual sound quality across both tasks. Furthermore, dCoNNear not only accurately emulated all non-DNN-based biophysical auditory models but also achieved comparable compensation performance to the prior CoNNear in hearing-aid processing and outperformed it in speech enhancement.
Overall, the dCoNNear-based closed-loop framework holds great promise for advancing audio technologies, including hearing aids and speech enhancement with high sound quality.

\bibliographystyle{IEEEtran}
\bibliography{main}

@inproceedings{donahue2018adversarial,
  title={Adversarial Audio Synthesis},
  author={Donahue, Chris and McAuley, Julian and Puckette, Miller},
  booktitle={International Conference on Learning Representations},
  year={2018}
}

@ARTICLE{SEQualityNet,
  author={Fu, Szu-Wei and Liao, Chien-Feng and Tsao, Yu},
  journal={IEEE Signal Processing Letters}, 
  title={Learning With Learned Loss Function: Speech Enhancement With Quality-Net to Improve Perceptual Evaluation of Speech Quality}, 
  year={2020},
  volume={27},
  number={},
  pages={26-30},
  doi={10.1109/LSP.2019.2953810}}

@inproceedings{germain2019speech,
  title={Speech Denoising with Deep Feature Losses},
  author={Germain, Francois and Chen, Qifeng and Koltun, Vladlen},
  booktitle={Proceedings of the Annual Conference of the International Speech Communication Association, INTERSPEECH},
  year={2019}
}

@inproceedings{zhang2019making,
  title={Making convolutional networks shift-invariant again},
  author={Zhang, Richard},
  booktitle={International conference on machine learning},
  pages={7324--7334},
  year={2019},
  organization={PMLR}
}

@article{karras2021alias,
  title={Alias-free generative adversarial networks},
  author={Karras, Tero and Aittala, Miika and Laine, Samuli and H{\"a}rk{\"o}nen, Erik and Hellsten, Janne and Lehtinen, Jaakko and Aila, Timo},
  journal={Advances in neural information processing systems},
  volume={34},
  pages={852--863},
  year={2021}
}

@article{kong2020hifi,
  title={Hifi-gan: Generative adversarial networks for efficient and high fidelity speech synthesis},
  author={Kong, Jungil and Kim, Jaehyeon and Bae, Jaekyoung},
  journal={Advances in neural information processing systems},
  volume={33},
  pages={17022--17033},
  year={2020}
}

@inproceedings{bak2023avocodo,
  title={Avocodo: Generative adversarial network for artifact-free vocoder},
  author={Bak, Taejun and Lee, Junmo and Bae, Hanbin and Yang, Jinhyeok and Bae, Jae-Sung and Joo, Young-Sun},
  booktitle={Proceedings of the AAAI Conference on Artificial Intelligence},
  volume={37},
  number={11},
  pages={12562--12570},
  year={2023}
}

@article{xu2021deep,
  title={Deep noise suppression with non-intrusive pesqnet supervision enabling the use of real training data},
  author={Xu, Ziyi and Strake, Maximilian and Fingscheidt, Tim},
  journal={arXiv preprint arXiv:2103.17088},
  year={2021}
}

@inproceedings{fu2022metricgan,
  title={MetricGAN-U: Unsupervised speech enhancement/dereverberation based only on noisy/reverberated speech},
  author={Fu, Szu-Wei and Yu, Cheng and Hung, Kuo-Hsuan and Ravanelli, Mirco and Tsao, Yu},
  booktitle={ICASSP 2022-2022 IEEE International Conference on Acoustics, Speech and Signal Processing (ICASSP)},
  pages={7412--7416},
  year={2022},
  organization={IEEE}
}

@article{shang2023analysis,
  title={Analysis and solution to aliasing artifacts in neural waveform generation models},
  author={Shang, Zengqiang and Zhang, Haozhe and Zhang, Pengyuan and Wang, Li and Li, Ta},
  journal={Applied Acoustics},
  volume={203},
  pages={109183},
  year={2023},
  publisher={Elsevier}
}

@article{drakopoulos2023neural,
  title={A neural-network framework for the design of individualised hearing-loss compensation},
  author={Drakopoulos, Fotios and Verhulst, Sarah},
  journal={IEEE/ACM Transactions on Audio, Speech, and Language Processing},
  year={2023},
  publisher={IEEE}
}

@inproceedings{pons2021upsampling,
  title={Upsampling artifacts in neural audio synthesis},
  author={Pons, Jordi and Pascual, Santiago and Cengarle, Giulio and Serr{\`a}, Joan},
  booktitle={ICASSP 2021-2021 IEEE International Conference on Acoustics, Speech and Signal Processing (ICASSP)},
  pages={3005--3009},
  year={2021},
  organization={IEEE}
}

@article{odena2016deconvolution,
  title={Deconvolution and checkerboard artifacts},
  author={Odena, Augustus and Dumoulin, Vincent and Olah, Chris},
  journal={Distill},
  volume={1},
  number={10},
  pages={e3},
  year={2016}
}

@inproceedings{zhang2018DFSMN,
  title={Deep-FSMN for large vocabulary continuous speech recognition},
  author={Zhang, Shiliang and Lei, Ming and Yan, Zhijie and Dai, Lirong},
  booktitle={2018 IEEE International Conference on Acoustics, Speech and Signal Processing (ICASSP)},
  pages={5869--5873},
  year={2018},
  organization={IEEE}
}

@inproceedings{lea2016temporal,
  title={Temporal convolutional networks: A unified approach to action segmentation},
  author={Lea, Colin and Vidal, Rene and Reiter, Austin and Hager, Gregory D},
  booktitle={Computer Vision--ECCV 2016 Workshops: Amsterdam, The Netherlands, October 8-10 and 15-16, 2016, Proceedings, Part III 14},
  pages={47--54},
  year={2016},
  organization={Springer}
}

@article{baby2021convolutional,
  title={A convolutional neural-network model of human cochlear mechanics and filter tuning for real-time applications},
  author={Baby, Deepak and Van Den Broucke, Arthur and Verhulst, Sarah},
  journal={Nature machine intelligence},
  volume={3},
  number={2},
  pages={134--143},
  year={2021},
  publisher={Nature Publishing Group UK London}
}

@article{drakopoulos2021convolutional,
  title={A convolutional neural-network framework for modelling auditory sensory cells and synapses},
  author={Drakopoulos, Fotios and Baby, Deepak and Verhulst, Sarah},
  journal={Communications Biology},
  volume={4},
  number={1},
  pages={827},
  year={2021},
  publisher={Nature Publishing Group UK London}
}

@article{stoller2018waveUNet,
  title={Wave-u-net: A multi-scale neural network for end-to-end audio source separation},
  author={Stoller, Daniel and Ewert, Sebastian and Dixon, Simon},
  journal={arXiv preprint arXiv:1806.03185},
  year={2018}
}

@article{zaidi2022daft,
  title={Daft-Exprt: Cross-Speaker Prosody Transfer on Any Text for Expressive Speech Synthesis},
  author={Za{\i}di, Julian and Seut{\'e}, Hugo and van Niekerk12, Benjamin and Carbonneau, Marc-Andr{\'e}},
  year={2022}
}

@article{abdel2014convolutional,
  title={Convolutional neural networks for speech recognition},
  author={Abdel-Hamid, Ossama and Mohamed, Abdel-rahman and Jiang, Hui and Deng, Li and Penn, Gerald and Yu, Dong},
  journal={IEEE/ACM Transactions on audio, speech, and language processing},
  volume={22},
  number={10},
  pages={1533--1545},
  year={2014},
  publisher={IEEE}
}

@article{park2017ConvSE,
  title={A Fully Convolutional Neural Network for Speech Enhancement},
  author={Park, Se Rim and Lee, Jin Won},
  journal={Evaluation},
  volume={10},
  pages={5},
  year={2017}
}

@article{verhulst2012TL,
  title={Nonlinear time-domain cochlear model for transient stimulation and human otoacoustic emission},
  author={Verhulst, Sarah and Dau, Torsten and Shera, Christopher A},
  journal={The Journal of the Acoustical Society of America},
  volume={132},
  number={6},
  pages={3842--3848},
  year={2012},
  publisher={AIP Publishing}
}

@article{verhulst2018computational,
  title={Computational modeling of the human auditory periphery: Auditory-nerve responses, evoked potentials and hearing loss},
  author={Verhulst, Sarah and Altoe, Alessandro and Vasilkov, Viacheslav},
  journal={Hearing research},
  volume={360},
  pages={55--75},
  year={2018},
  publisher={Elsevier}
}

@article{garofolo1993TIMIT,
  title={DARPA TIMIT acoustic-phonetic continous speech corpus CD-ROM. NIST speech disc 1-1.1},
  author={Garofolo, John S and Lamel, Lori F and Fisher, William M and Fiscus, Jonathan G and Pallett, David S},
  journal={NASA STI/Recon technical report n},
  volume={93},
  pages={27403},
  year={1993}
}

@article{altoe2017ihc,
  title={Model-based estimation of the frequency tuning of the inner-hair-cell stereocilia from neural tuning curves},
  author={Alto{\`e}, Alessandro and Pulkki, Ville and Verhulst, Sarah},
  journal={The Journal of the Acoustical Society of America},
  volume={141},
  number={6},
  pages={4438--4451},
  year={2017},
  publisher={AIP Publishing}
}

@article{altoe2018ANF,
  title={The effects of the activation of the inner-hair-cell basolateral K+ channels on auditory nerve responses},
  author={Altoe, Alessandro and Pulkki, Ville and Verhulst, Sarah},
  journal={Hearing research},
  volume={364},
  pages={68--80},
  year={2018},
  publisher={Elsevier}
}

@article{kingma2014adam,
  title={Adam: A method for stochastic optimization},
  author={Kingma, Diederik P and Ba, Jimmy},
  journal={arXiv preprint arXiv:1412.6980},
  year={2014}
}

@article{keshishzadeh2021towards,
  title={Towards personalized auditory models: Predicting individual sensorineural hearing-loss profiles from recorded human auditory physiology},
  author={Keshishzadeh, Sarineh and Garrett, Markus and Verhulst, Sarah},
  journal={Trends in Hearing},
  volume={25},
  pages={2331216520988406},
  year={2021},
  publisher={SAGE Publications Sage CA: Los Angeles, CA}
}

@article{greenwood1990cochlear,
  title={A cochlear frequency-position function for several species—29 years later},
  author={Greenwood, Donald D},
  journal={The Journal of the Acoustical Society of America},
  volume={87},
  number={6},
  pages={2592--2605},
  year={1990},
  publisher={Acoustical Society of America}
}

@inproceedings{van2020hearing,
  title={Hearing-impaired bio-inspired cochlear models for real-time auditory applications},
  author={Van Den Broucke, Arthur and Baby, Deepak and Verhulst, Sarah},
  booktitle={21st Annual Conference of the International Speech Communication Association (INTERSPEECH 2020)},
  pages={2842--2846},
  year={2020},
  organization={International Speech Communication Association (ISCA)}
}

@article{falk2010SRMR,
  title={A non-intrusive quality and intelligibility measure of reverberant and dereverberated speech},
  author={Falk, Tiago H and Zheng, Chenxi and Chan, Wai-Yip},
  journal={IEEE Transactions on Audio, Speech, and Language Processing},
  volume={18},
  number={7},
  pages={1766--1774},
  year={2010},
  publisher={IEEE}
}

@article{shmilovitz2005THD,
  title={On the definition of total harmonic distortion and its effect on measurement interpretation},
  author={Shmilovitz, Doron},
  journal={IEEE Transactions on Power delivery},
  volume={20},
  number={1},
  pages={526--528},
  year={2005},
  publisher={IEEE}
}

@inproceedings{fonseca2017freesound,
  title={Freesound datasets: a platform for the creation of open audio datasets},
  author={Fonseca, Eduardo and Pons Puig, Jordi and Favory, Xavier and Font Corbera, Frederic and Bogdanov, Dmitry and Ferraro, Andres and Oramas, Sergio and Porter, Alastair and Serra, Xavier},
  booktitle={Hu X, Cunningham SJ, Turnbull D, Duan Z, editors. Proceedings of the 18th ISMIR Conference; 2017 oct 23-27; Suzhou, China.[Canada]: International Society for Music Information Retrieval; 2017. p. 486-93.},
  year={2017},
  organization={International Society for Music Information Retrieval (ISMIR)}
}

@article{drakopoulos2022model,
  title={Model-based hearing-enhancement strategies for cochlear synaptopathy pathologies},
  author={Drakopoulos, Fotios and Vasilkov, Viacheslav and Vecchi, Alejandro Osses and Wartenberg, Tijmen and Verhulst, Sarah},
  journal={Hearing Research},
  volume={424},
  pages={108569},
  year={2022},
  publisher={Elsevier}
}

@inproceedings{fma_dataset,
  title = {{FMA}: A Dataset for Music Analysis},
  author = {Defferrard, Micha\"el and Benzi, Kirell and Vandergheynst, Pierre and Bresson, Xavier},
  booktitle = {18th International Society for Music Information Retrieval Conference (ISMIR)},
  year = {2017},
  archiveprefix = {arXiv},
  eprint = {1612.01840},
  url = {https://arxiv.org/abs/1612.01840},
}

@ARTICLE{conv-tasnet,
  author={Luo, Yi and Mesgarani, Nima},
  journal={IEEE/ACM Transactions on Audio, Speech, and Language Processing}, 
  title={Conv-TasNet: Surpassing Ideal Time–Frequency Magnitude Masking for Speech Separation}, 
  year={2019},
  volume={27},
  number={8},
  pages={1256-1266},
  doi={10.1109/TASLP.2019.2915167}}

@inproceedings{drakopoulos2023dnn,
  title={A DNN-based hearing-aid strategy for real-time processing: One size fits all},
  author={Drakopoulos, Fotios and Van Den Broucke, Arthur and Verhulst, Sarah},
  booktitle={ICASSP 2023-2023 IEEE International Conference on Acoustics, Speech and Signal Processing (ICASSP)},
  pages={1--5},
  year={2023},
  organization={IEEE}
}

@inproceedings{drakopoulos2022differentiable,
  title={A differentiable optimisation framework for the design of individualised DNN-based hearing-aid strategies},
  author={Drakopoulos, Fotios and Verhulst, Sarah},
  booktitle={ICASSP 2022-2022 IEEE International Conference on Acoustics, Speech and Signal Processing (ICASSP)},
  pages={351--355},
  year={2022},
  organization={IEEE}
}

@article{pesq2001perceptual,
  title={Perceptual evaluation of speech quality (PESQ): An objective method for end-to-end speech quality assessment of narrow-band telephone networks and speech codecs},
  author={Recommendation, ITU-T},
  journal={Rec. ITU-T P. 862},
  year={2001}
}

@inproceedings{reddy2021dnsmos,
  title={DNSMOS: A non-intrusive perceptual objective speech quality metric to evaluate noise suppressors},
  author={Reddy, Chandan KA and Gopal, Vishak and Cutler, Ross},
  booktitle={ICASSP 2021-2021 IEEE International Conference on Acoustics, Speech and Signal Processing (ICASSP)},
  pages={6493--6497},
  year={2021},
  organization={IEEE}
}

@article{zen2019libritts,
  title={Libritts: A corpus derived from librispeech for text-to-speech},
  author={Zen, Heiga and Dang, Viet and Clark, Rob and Zhang, Yu and Weiss, Ron J and Jia, Ye and Chen, Zhifeng and Wu, Yonghui},
  journal={arXiv preprint arXiv:1904.02882},
  year={2019}
}

@inproceedings{rix2001perceptual,
  title={Perceptual evaluation of speech quality (PESQ)-a new method for speech quality assessment of telephone networks and codecs},
  author={Rix, Antony W and Beerends, John G and Hollier, Michael P and Hekstra, Andries P},
  booktitle={2001 IEEE international conference on acoustics, speech, and signal processing. Proceedings (Cat. No. 01CH37221)},
  volume={2},
  pages={749--752},
  year={2001},
  organization={IEEE}
}

@article{jensen2016algorithm,
  title={An algorithm for predicting the intelligibility of speech masked by modulated noise maskers},
  author={Jensen, Jesper and Taal, Cees H},
  journal={IEEE/ACM Transactions on Audio, Speech, and Language Processing},
  volume={24},
  number={11},
  pages={2009--2022},
  year={2016},
  publisher={IEEE}
}

@article{reddy2021interspeech,
  title={Interspeech 2021 deep noise suppression challenge},
  author={Reddy, Chandan KA and Dubey, Harishchandra and Koishida, Kazuhito and Nair, Arun and Gopal, Vishak and Cutler, Ross and Braun, Sebastian and Gamper, Hannes and Aichner, Robert and Srinivasan, Sriram},
  journal={arXiv preprint arXiv:2101.01902},
  year={2021}}

@inproceedings{gemmeke2017audio,
  title={Audio set: An ontology and human-labeled dataset for audio events},
  author={Gemmeke, Jort F and Ellis, Daniel PW and Freedman, Dylan and Jansen, Aren and Lawrence, Wade and Moore, R Channing and Plakal, Manoj and Ritter, Marvin},
  booktitle={2017 IEEE international conference on acoustics, speech and signal processing (ICASSP)},
  pages={776--780},
  year={2017},
  organization={IEEE}
}

\end{document}